\documentclass[11pt,a4paper]{article}
\pdfoutput=1
\usepackage[usenames,dvipsnames]{xcolor}
\usepackage{jheppubx}
\usepackage{enumitem}
\usepackage{graphicx}
\usepackage{dcolumn}
\usepackage{bm,psfrag}
\usepackage{amsmath}
\usepackage{cleveref}
\usepackage{subfigure}
\usepackage{float}
\usepackage[sf,isu]{caption}
\usepackage{bbm}

\hypersetup{pdftitle={On thermality of CFT eigenstates},    
	pdfauthor={ddp}, 
	colorlinks=true, 
	linkcolor=blue, 
	citecolor=green!39!black!,
	filecolor=OliveGreen, 
	urlcolor=RoyalBlue}

 \usepackage{overpic}
\def\bra#1{{\langle}#1|}
\def\ket#1{|#1\rangle}
\def\Tr{{{\rm Tr}}}
\def\CO{{\cal O}}

\newcommand{\be}{\begin{equation}}
\newcommand{\ee}{\end{equation}}

\usepackage{chngcntr}
\counterwithout{equation}{section} 

\makeatletter
\g@addto@macro\bfseries{\boldmath}
\makeatother

\newcommand{\bea}{\begin{eqnarray}}
\newcommand{\eea}{\end{eqnarray}}
\newcommand{\ba}{\begin{eqnarray}}
\newcommand{\ea}{\end{eqnarray}}
\newcommand{\nn}{\nonumber \\}

\newcommand{\beq}{\begin{equation}}
\newcommand{\eeq}{\end{equation}}
\newcommand{\beqa}{\begin{eqnarray}}
\newcommand{\eeqa}{\end{eqnarray}}
\newcommand{\beqar}{\begin{eqnarray*}}
\newcommand{\eeqar}{\end{eqnarray*}}

\newcommand{\eg}{{\it e.g.,}\ }
\newcommand{\ie}{{\it i.e.,}\ }

\def\tr{\rm tr}

\def\t6 {T_\mt{D6}}

\newcommand{\mt}[1]{\textrm{\tiny #1}}

\newcommand{\cA}{{\cal A}}

\def\cale         {{\cal E}}

\def\ee           {{\rm e}}

\def\tr           {\mathop{\rm Tr}}

\def\sqr#1#2{{\vcenter{\vbox{\hrule height.#2pt
 \hbox{\vrule width.#2pt height#1pt \kern#1pt
 \vrule width.#2pt}\hrule height.#2pt}}}}

\usepackage{lipsum} 

\def\ee{\cale}

\def\aa1{\phi}
\def\cc1{\psi}

\def\T{\mathcal{T}}

\def\ra{\Longrightarrow}
\def\vev#1{\langle #1 \rangle}

\def\bt{\widetilde{\mathbbm{B}}}

\def\la{\langle}
\def\ra{\rangle}

\def\nnn{\nonumber}

\def\vev#1{\langle{#1}\rangle}

\def\opr{{\mathbbm{O}}}

\def\nj{{ \vphantom{(j)}}}

\newcommand{\normord}[1]{\vcentcolon\mathrel{#1}\vcentcolon}
\providecommand{\vcentcolon}{\mathrel{\mathop{:}}}

\begin{document}

\title{On thermality of CFT eigenstates}

\author[1]{Pallab Basu}
\author[2]{\!, Diptarka Das}
\author[3]{\!, Shouvik Datta}
\author[2]{and Sridip Pal}

\affiliation[1]{\vspace{.3cm} International Center for Theoretical Sciences-TIFR,\\ Shivakote, Hesaraghatta Hobli, \\ Bengaluru North 560089, India. \\ \vspace{-.3cm}
}
\affiliation[2]{Department of Physics,\\ University of California San Diego,\\
	La Jolla, CA 92093, USA.\\ \vspace{-.3cm}
}
\affiliation[3]{Institut f\"{u}r Theoretische Physik,\\ Eidgen\"{o}ssische Technische Hochschule Z\"{u}rich, \\ 
	8093 Z\"{u}rich, Switzerland.\\  \vspace{0.2cm}
}


\emailAdd{pallab.basu@icts.res.in}
\emailAdd{diptarka@physics.ucsd.edu}
\emailAdd{\\ shouvik@itp.phys.ethz.ch}
\emailAdd{srpal@ucsd.edu}

\abstract{The Eigenstate Thermalization Hypothesis (ETH) provides a way to understand how an isolated quantum mechanical system can be approximated by a thermal density matrix. We find a class of operators in (1+1)-$d$ conformal field theories, consisting of quasi-primaries of the identity module, which satisfy the hypothesis only at the leading order in large central charge. In the context of subsystem ETH, this plays a role in the deviation of the reduced density matrices, corresponding to a finite energy density eigenstate from its hypothesized thermal approximation. The universal deviation in terms of the square of the trace-square distance goes as the 8th power of the subsystem fraction and is suppressed by powers of inverse central charge ($c$). Furthermore, the non-universal deviations from subsystem ETH are found to be proportional to the heavy-light-heavy structure constants which are typically exponentially suppressed in $\sqrt{h/c}$, where $h$ is the conformal scaling dimension of the finite energy density state. We also examine the effects of the leading finite size corrections.}

\hypersetup{urlcolor=Gray!75!black}
\maketitle
\hypersetup{urlcolor=RoyalBlue}

\section{Introduction and summary of results}
From cold atoms to quantum gravity, the question of how an isolated and interacting system reaches thermal equilibrium under unitary time evolution, is of central importance in quantum statistical mechanics. A sharper notion of thermalization arises from considering subsystems (say, $\cA$). If the subsystem thermalizes, the rest of the system (for $\bar{\cA}$ sufficiently large) serves as heat bath or a reservoir which enables the subsystem to equilibrate. In other words, the reduced density matrix, $\rho_\cA(t)$ approaches the thermal reduced density  matrix $\rho_\cA^\beta$ at late times. Moreover, this expectation of thermalization comes the  observation  that, a typical finite energy density state, after being evolved for a sufficiently long time gives prediction according to the rules of equilibrium statistical mechanics. 

The eigenstate thermalization hypothesis (ETH) makes this precise by positing that thermalization holds for every individual energy eigenstate $\ket{\psi}$ of the Hamiltonian \cite{Deutsch:1991, Srednicki:1994,rigol2008thermalization,popescu2006entanglement,nandkishore2015many}. When the initial state is an energy eigenstate, the unitary time evolution is trivial, $\rho(t)=\rho(0)=\ket{\psi}\bra{\psi}$. This implies that if a system thermalizes (for all choices of the initial state), all eigenstates of the Hamiltonian are thermal. As a consequence, the expectation values of operators evaluated at thermal equilibrium should equal those evaluated in the eigenstate, $\psi$.
\beq
\bra{\psi } \opr \ket{\psi} = \frac{ \Tr \left(\opr\, e^{-\beta H} \right)}{\Tr \left(e^{-\beta H }\right)}, \label{thermalization}
\eeq 
In the above equation, which we call the $global$ ETH condition for operator $\opr$, the inverse temperature $\beta$ is chosen by fixing the energy eigenvalue, $\bra{\psi} H \ket{\psi} = E_{\psi} =\la H \ra_\beta$. In terms of the reduced density matrix $\rho_\cA^\psi$, \eqref{thermalization} is equivalent to,
\bea\label{thermalization-2}
\Tr_\cA ( \opr \rho_\cA^\psi )  = \Tr_A ( \opr \rho^{\beta}_\cA ) \qquad \qquad \text{where \ \ } \rho_\cA^{\beta} = \frac{ \Tr_{\bar{\cA}} (e^{ -\beta H })}{\Tr (e^{\beta H}) }
\eea
 Clearly if $all$ local and non-local operators in region $A$ satisfy \eqref{thermalization} then the above equation implies that $\rho_\cA^\psi = \rho_\cA^{\beta}$, which is called the subsystem ETH. If the subsystem ETH is true, the entanglement entropies of the reduced density matrices should also be the same. A large number of verifications of the hypothesis, involving the numerical extraction of eigenstates by exact diagonalization, has been carried out in various lattice models \cite{rigol2008thermalization,huse2014,Garrison:2015lva,alba2015eigenstate,Nandy:2016fwv}. It has also been proved for quantum systems having a classical chaotic limit \cite{berry1977regular,Srednicki:1994}. A natural diagnostic to quantify the equivalence of the reduced density matrices is to define a notion of distance. Two useful distance functions are the trace two norm $||\cdot||_{2}$ and trace one norm $||\cdot||_{1}$ defined as,  
 \beq\label{definitions}
 ||G||_{1}= \Tr \sqrt{G^2} =  \sum_{i}|\lambda_{i}|  \qquad, \qquad ||G||_{2}= \sqrt{\Tr G^2} = \left({\sum_{i}\lambda_{i}^{2}}\right)^{1/2}.
 \eeq
 where $\lambda_{i}$ are eigenvalues of $G$. In the present scenario, we are interested in $G = \rho_\cA^\psi - \rho_\cA^{th}$. The two-norm or the trace-square distance (TSD) shall play a central role in this paper. Distances of the above kind also serve as suitable order parameters (in the space of theories/initial-states) to detect transitions between thermalization and many-body localized phase \cite{nandkishore2015many}. 

It has been recently shown in \cite{Garrison:2015lva} that  if ETH holds good, then it allows one to use a single eigenstate to extract properties of the Hamiltonian at arbitrary temperatures. 
\begin{enumerate}[leftmargin=*]
\item \label{conj1} It is conjectured and supported by numerical evidence in \cite{Garrison:2015lva} that \eqref{thermalization-2} holds true for all operators  \textit{only as long as} $V_\cA /V \ll 1$; here $V_\cA$ is the volume of subsystem $\cA$ and $V$ is the total volume.\footnote{More precisely this is the limit, $V_\cA/V \rightarrow 0$ as $V\rightarrow \infty$. There is also  numerical evidence in different $1+1$ dimensional lattice models that there are a large set of operators for which \eqref{thermalization-2} is true for $1/2> f^*>V_A/V > 0$, for some $O(1)$ number $f^{*}$ \cite{Garrison:2015lva}. } Hence, subsystem ETH holds true only in this limit. This is equivalent to the statement that subsystem ETH, $\rho_{\mathcal{A}}^\psi = \rho_{\mathcal{A}}^\beta$ is true for $V_\cA /V \rightarrow 0$.

The reason why this fails for finite $V_\cA/V$ is because there are some operators for which \cref{thermalization} is not true. One common example is the energy variance operator, $(H_{\mathcal{A}}- \vev{H_{\mathcal{A}}} )^2$. 

\item \label{conj2} More precisely, \cite{Garrison:2015lva} also conjectures that for $V_\cA/V > 0$ the set of operators spanned by $H_\cA^n$ for $n>1$ do not satisfy the ETH relation \eqref{thermalization-2}. Here $H_\cA$ is the Hamiltonian restricted to the subsystem $\cA$. 

\end{enumerate}

\begin{enumerate}[leftmargin=*]
\setcounter{enumi}{2}
\item\label{conj3} For the one norm, numerics in \cite{Garrison:2015lva} for the $1+1$ dimensional quantum spin 1/2 model with both longitudinal and transverse field turned on, suggest that there is a scaling of this distance with the subsystem fraction. In the regime the subsystem size ($\ell$) is much smaller than the system size ($L$) and $\beta$ smaller than all scales, then the trace norm distance tends to zero at least linearly with $1/L$. Nonetheless, there is no analytical evidence of what this scaling behavior should be.
\end{enumerate}

An interesting arena to explore the intricacies of ETH and the related conjectures above is that of 2D conformal field theories.
Besides describing the critical point of statistical systems and being duals to 3D gravity, 2D CFTs are tractable analytically owing to the infinite dimensional Virasoro symmetry. 
For generic states in the CFT, the equilibrium ensemble is the Generalized Gibbs Ensemble (GGE) \cite{Bernard:2016nci}. However for small subsystems, and for finite energy density states we shall find the Gibbs Ensemble to be a good approximant. 
We provide analytical support for conjectures \ref{conj1} and \ref{conj2} for 2D CFTs, within the regime of validity of our calculations. We also derive scaling of the trace-square distance \eqref{definitions} with $\ell/L$ that lends analytic support to the observed scaling of the one-norm  \ref{conj3}. 

\subsection*{Subsystem ETH in 2D CFTs} 

Recently it has been shown in \cite{Lashkari:2016vgj}  that the subsystem ETH is related to the a local form of ETH which states off-diagonal matrix elements of energy eigenstates are entropically suppressed. Moreover, for CFTs another measure for equivalence for density matrices, is $I(t) \ = \  \Tr_\cA ( \rho_\cA^{\alpha}(t) \rho_\cA^{\beta} )/\big[\Tr_\cA ( \rho_\cA^{\alpha}(t))^2\, \Tr_A (\rho_\cA^{\beta})^2  \big]^{1/2}$. This has been utilized in the context of sudden quantum critical quenches  in \cite{Cardy:2014rqa}, and generalized to the case of CFTs with additional conserved charges in \cite{ Mandal:2015jla}. However, in this case the initial finite energy density state is  \textit{not} an  eigenstate, but rather approximated by the conformal boundary state, upto irrelevant perturbations. {Studies on entanglement entropy and conformal blocks, revealing thermal aspects of CFT heavy states, can also be found in  \cite{Caputa:2014eta,Asplund:2014coa,Fitzpatrick:2015foa, Fitzpatrick:2015dlt, Fitzpatrick:2015zha, Dymarsky:2016aqv, Anous:2016kss,He:2017vyf}.}

 Our working model is that of a 1+1-dimensional CFT on a circle (of circumference $L$)  by considering a \textit{finite}
energy density \textit{eigenstate}, $\ket{\psi}$. We shall calculate the square of the trace square distance, $\T = || \rho_\cA^\psi - \rho_\cA^{\beta} ||_2^2$ in the regime of short subsystem sizes, $\ell\ll\beta\ll L$ (this includes the case with small but finite subsystem fraction). The CFT  eigenstates $\ket{\psi}$ which we consider are created by insertion of a heavy scalar primary of conformal dimension $h\gg 1$, at Euclidean time $\tau=-\infty$ of the cylinder. 
It has been shown   that, at large central charge, the entanglement entropy matches with the thermal entanglement entropy \cite{Asplund:2014coa}. However, the higher Renyi entropies do not agree at arbitrary values of the central charge. This indicates that subsystem ETH is, generically, not satisfied for CFT.
Upon explicitly calculating the TSD, we find a number of features which are summarized below. 

\subsection*{Summary of results}

 In the thermodynamic limit, we show a version of the conjecture \ref{conj2} for quasi-primaries of the conformal family of the identity of level $2n$ for $n>1$ for finite $c$. The leading deviation from the global ETH condition \eqref{thermalization} comes from the quasi-primary containing the term $\normord{T^n}$. This deviation goes as $1/{c}$. Therefore, for large central charge ETH holds for these operators. For the sake of clarity, we must mention, here we are dealing with a double scaling limit, first we take the thermodynamic limit and let $L\rightarrow\infty$, which makes $h \sim L^{2}\gg1$ and then we take large central charge limit. For primary operators $\CO$ and its descendants the deviation is typically zero in this limit.

 The trace square distance, $\T$, is computed in the short interval expansion \ie order-by-order in powers of $\ell/L$. The interval size $\ell$ is much smaller than all length scales in problem. We find that that the leading behaviour is  $(\ell/L)^{\gamma}$ in the $\ell/L\rightarrow 0$ limit. This confirms conjecture \ref{conj1}. For a CFT with no light scalar primaries of dimension $\Delta_{\phi} < 4$, we have $\gamma = 8$. This is a universal feature and comes from the quasi-primary of the identity module at level 4. When the lightest primary $\phi$ is such that $\Delta_{\phi} < 4$, we have $\gamma = 2\Delta_{\phi}$.

As a main result of our paper, we find that the trace square distance can be written as
\bea\label{result}
\T \ &=& \ \sum_{\Theta, \Theta'} C_{\Theta, \Theta'}  \bigg( \bra{\psi} \Theta \ket{\psi} - \vev{\Theta}_{\rm thermal} \bigg)  \bigg( \bra{\psi} \Theta' \ket{\psi} - \vev{\Theta'}_{\rm thermal} \bigg).
\eea
Here, the sum is over all quasi-primaries ($\Theta, \Theta'$). The factors in the two parentheses are precisely the deviations from global ETH, for the operators $\Theta$ and $\Theta'$. \footnote{For the term in the sum with $\Theta = T$ (the stress tensor), the difference in the parentheses is zero, by the very definition of the temperature of the thermal ensemble, $\beta$ which is fixed in terms of $h$ and $c$, see equations \eqref{beta-def},\eqref{beta1}.} Here, $C_{\Theta,\Theta'}$ is determined by using the orthogonality of the quasi-primaries on the plane. Thus we see that the ETH deviations of $\Theta, \Theta'$ have a  direct contribution to the trace square distance!  
Next we look separately at the universal and non-universal contributions to $\T$. In the limit, $L\rightarrow \infty$, with $h\sim L^2$ and next taking $c$ large we find at the leading order, 
\begin{align}
\hat\T &= \hat\T_{\text{univ}} +\hat\T_{\text{non-univ}}. \nonumber \\
\hat\T_{\text{univ}} &=   \frac{1}{c^2}\ \sum_{n\geq2 } \alpha_{n} \bigg(\frac{h \ell^2}{L^2}\bigg)^{2 n}. \label{summary}\\ 
\hat\T_{\text{non-univ}} &= {\sum_{\phi \notin {\mathcal{V}_{\mathbb{I}}}} \alpha_\phi \bigg(\frac{\ell}{L}\bigg)^{2 \Delta_\phi } c_{\chi\phi \chi}^2\ {\rm O}(e^{-8\pi \Delta_\chi \sqrt{h/c}})}. \nonumber
\end{align}
In writing $\hat \T$, we have normalized $\T$ by the second Renyi entropy of the CFT vacuum on a plane to remove UV divergences. Here, $\alpha_{n}, \alpha_\phi$ are numbers independent of $h$ and $c$. In the above equation, the contribution to $\hat \T_{\text{univ}}$ is from quasi-primary operators, in the identity module from levels $n\geq 4$, which contain, $\normord{T^n}$. We see in the large $c$ limit, these are suppressed by $1/c^2$. Away from the large $c$ limit, this particular class of universal operators violate the global ETH \eqref{thermalization} and hence by \eqref{result} directly contributes to $\T$. The behavior of the non-universal piece is  typically  exponentially suppressed. Here $\Delta_{\chi}$ is the smallest dimension among the operators with non-vanishing expectation value $c_{\chi \phi \chi}$.

Our calculations are done in the thermodynamic limit, ignoring finite size effects. Conformal symmetry fixes the leading finite size corrections. Taking this into account we shall see that the relationship between $\beta$ and $h$ gets exponentially small corrections. However, for the trace square distance, $\T$, these finite size corrections to $\beta$ are super-exponentially suppressed. There are also more direct exponentially small finite size corrections to $\vev{\Theta}_{\text{thermal}}$ which go as powers of $e^{-L/\beta} \sim e^{-\sqrt{h/c}}$. These do not change the behaviors of \eqref{summary} above.
 
 This paper is organized as follows. In \S\ref{dtp}, we describe our set-up and elucidate on the validity of ETH at large central charge.  We compute the trace two norm squared order by order in powers of subsystem size, explicitly evaluating universal contributions coming from the identity module as well as non-universal ones from the primary operators in  \S\ref{section:tracedistance}. In \S\ref{section:deviationfromETH}, we evaluate the deviation from ETH for both the universal and the non-universal sectors discussing the finite size effects. We conclude in \S\ref{sec:last} with discussions and future avenues. Some details of the calculations are presented to the appendices \ref{app-b} and \ref{app-c}. As a by-product, we also present the second Renyi entropy for the finite energy density eigenstate till quartic order in subsystem fraction in appendix \ref{app-d}.

\section{Testing ETH for operators in CFT${}_2$}\label{dtp}
We work with a CFT on a circle of circumference $L$. The conformal transformation, 
$z = e^{-\frac{2\pi i}{L} w }$, maps the cylinder onto the plane. Here $w=x + i\tau$ is the euclidean coordinate on the cylinder, while $z$ is the coordinate on the plane. We can use the map to create a spinless homogenous energy eigenstate, by inserting 
a local primary $\psi$ of conformal weight $(h,h)$ at ($z=\bar{z} = 0$) on the plane as considered in \cite{Sarosi:2016oks, Asplund:2014coa}. In terms of the coordinates on the cylinder, the primary is located at $\tau = -\infty$. Using the state-operator correspondence the eigenstate is given by
\beq
\ket{\psi} \ = \ \psi(z_2=\bar{z}_2 = 0)\ket{0},\label{state}
\eeq
while its dagger is, 
\beq
\bra{\psi }  =\lim\limits_{z_1, \bar{z}_1 \rightarrow \infty} z_1^{2h} \bar z_1^{2\bar h}\bra{0} \psi^\dagger(z_1,\bar{z}_1).  
\eeq 
This ensures that the states are properly normalized i.e. $\bra{\psi}\psi\rangle = 1$.  The energy density goes as $\frac{2h}{L^{2}}$. An eigenstate with finite energy density is obtained by considering a primary operator $\psi$ such that $h$ goes as $L^2$. In thermodynamic limit, $L$ is very large. We are therefore interested in the `heavy' limit $h \gg1$.

 \subsection*{Relating $\beta$ to the conformal weight } 
 The statement of global ETH \eqref{thermalization} approximates the energy eigenstate\footnote{This can be thought of as \textit{single eigenstate ensemble}, see \eg \cite{nandkishore2015many}. It is a special limit of the standard micro-canonical ensemble, where only one eigenstate lies within the energy window.} described by the density matrix $\rho^\psi = \ket{\psi}\bra{\psi}$ by a canonical ensemble described by the thermal density matrix $\rho^{\beta}  = \frac{1}{Z_\beta} e^{-\beta H}$. Here the inverse temperature $\beta$ is determined by setting the expectation value of the energy density to be equal, \ie 
 \beq \label{beta-def}
  \bra{\psi} T \ket{\psi} = \vev{T}_{\text{thermal}} \qquad 
  \implies \qquad
   \frac{ 4 \pi^2 h}{L^2}   -\frac{\pi^2 c}{6 L^2} \ = \  \frac{\pi^2 c}{6 \beta^2}.
 \eeq
The solution to the above is 
\beq
\beta = \beta_h = \frac{ L } {\sqrt{ 24h/{c} -1 } }. \label{beta1}
\eeq
It is to be noted that for large $h$, $\beta_h$ above is equal to the classical saddle value of inverse temperature appearing in the Cardy formula \cite{CARDY1986186}. It provides an alternative way to derive the Cardy formula for the density of states, if we use the thermodynamic relation $\left(\frac{\partial S}{\partial E}\right)_{L}= \beta_{h}$, as already noted in \cite{Fitzpatrick:2015foa}. Since finite energy density states require $h \gg1$ this implies that $\beta_h/ L \rightarrow 0$, which is consistent with working in the high temperature approximation. 

 \subsection{Global ETH at large central charge}\label{ETH violating operators}
 In this subsection, we consider quasi-primaries and compare their expectation values in the thermal state and in the eigenstate $\psi$. The commutation relation of a quasi-primary ($\Theta$) with the modes of the stress-tensor is
 \begin{align}
 [L_m,\Theta_n] = [(h_\Theta-1)m-n]\Theta_{m+n}.
 \nonumber
 \end{align}
where $h_\Theta$ is the conformal weight of $\Theta$. The zero modes, $\Theta_0$, therefore commute with $L_0$ and hence with the Hamiltonian. The quasi-primaries, therefore, form good observables to test eigenstate thermalization. Furthermore, these play a crucial role in our analysis of subsystem ETH in \S\ref{section:tracedistance}. Let us first focus on the quasi-primaries in the conformal family of the identity, which is universal for all CFTs. Among the level $2n$ quasi primaries, we call $\mathbbm{B}_n$, the ones which contains the term $\normord{T^{n}}$. Explicitly the first few operators of this kind are given by,
\bea
\mathbbm{B}_{1}&=&T,\nn
 \mathbbm{B}_{2}&=& \normord{TT}- \frac{3}{10} \partial^2 T, \nn
 \mathbbm{B}_{3}&=& \normord{T(\normord{TT})}  -\frac{9}{10}\normord{\partial^{2}TT}-\frac{1}{28}\partial^{4}T+\frac{93}{70c+29}\mathbbm{B}_2  .
\eea
The leading behaviour of $\langle\psi| \mathbbm{B}_{n}|\psi\rangle_{\text{cylinder}}$ as $L\rightarrow \infty$ comes from the piece $\normord{T^n}$ in $\mathbbm{B}_n$. This is easy to verify, using the recursive Ward identity for stress tensor insertions on plane
\begin{align*}
& \la T(z) T(z_1)\cdots T(z_M) \psi(y_1) \psi(y_2) \ra  \nonumber \\
& = \bigg[ \sum_{i=1}^2 \left( \frac{h}{(z-y_i)^2} + \frac{\partial_{y_i}}{(z-y_i)}\right) \la   T(z_1)\cdots T(z_M) \psi(y_1) \psi(y_2) \ra   \\
& \qquad + \sum_{j=1}^{M} \left({c/2 \over (z-z_i)^4} +  \frac{2}{(z-z_i)} + \frac{\partial_{z_i}}{(z-z_i)}\right)\la   T(z_1)\cdots T(z_M) \psi(y_1) \psi(y_2) \ra \bigg] .\nonumber
\end{align*}
It can be easily seen that the insertion of $M$ stress tensors in the correlator leads to a term having $h^M$. In the $h\gg 1$ limit, the domiant piece is 
\begin{align}
\la T(z_1)\cdots T(z_M) \psi(y_1) \psi(y_2) \ra \quad  \cong\quad  \frac{h^M}{|y_{12}|^{4h}}\ \prod_{i=1}^M\sum_{i=1,2} \frac{1}{(z_i-y_i)^2}.
\end{align}
{Clearly, in terms of the Laurent modes of the stress tensor, $\sum_m L_m/z^{m+2}$, the above contribution is purely from the term containing the zero modes, $\la\psi|(L_0)^M| \psi\ra$. 
 All the other terms in $\mathbbm{B}_n$ come with lower powers of $h$. Taking the limit of co-incident points for stress tensors  and conformal transforming to the cylinder, leads to 
 \begin{align}
 \vev{\psi| \mathbbm{B}_n | \psi} \cong \left(\frac{-4\pi^2h}{L^{2}}\right)^{n}
 \end{align}
Since $h$ scales as $L^2$ in the thermodynamic limit, the terms containing lower powers of $h$ and the terms coming from the Schwarzian goes to zero as ${L^{-\alpha}}$ with $\alpha >0$.  

On the other hand, $\langle \mathbbm{B}_{n} \rangle_{\text{thermal}}$ is given by $\dfrac{a_n}{\beta^{2n}}$, where $a_n$ represents the constant anomalous piece. For example,
\bea
a_1 = -\, \frac{\pi^2 c}{6}, \quad 
a_2 = \frac{ \pi^4 c( 5c + 22 )}{180}, \quad 
a_3 = -  \, \frac{ \pi^6 c(2c-1)(5c+22)(7c + 68)}{216(70c+29)}.
\eea
At leading order in central charge\footnote{We thank Kenneth Intriligator for pointing this out to us.}, the term with $c^n$ dominates for $a_n$ and
$
a_2 \cong (a_1)^2$ and $ a_3 \cong (a_1)^3 
$.
In general, it can be seen that, for quasi-primaries of the form $\mathbbm{B}_n \sim \, \normord{T^n}$, we have
\bea a_n =  \left(-\frac{\pi^2c}{6}\right)^n + \mathrm{O}(c^{n-1}) .
\label{a_n-behave}
\eea
This is generically true because of the transformation property of the stress tensor under conformal transformations
\begin{align}
T(w) = z'(w)^2 T(z) + \frac{c}{12} \lbrace z, w \rbrace .
\end{align}
While going from the plane to the thermal cylinder, the expectation value of the normal-ordered product of $n$ stress tensors, at large central charge, is dominated by $n$th power of anomalous piece involving the Schwarzian derivative ($s$). Note that for the conformal map from the plane to the thermal cylinder, $s$ is a constant.  

Therefore, in large $h$ limit, (with $\beta=\beta_{h}\approx L({{c}/{24h}})^{1/2}$ from \eqref{beta1}), we have,
\bea\label{compare}
\bigg(\langle\psi| \mathbbm{B}_{n}|\psi\rangle_{\text{cylinder}}-\langle \mathbbm{B}_{n} \rangle_{\text{thermal}}\bigg) \underset{L \rightarrow \infty}{=} \left(\frac{h}{L^{2}}\right)^{n}\left((-4\pi^2)^{n}-a_n\left(\frac{24}{c}\right)^{n}\right).
\eea
For $n=1$, the expression above is exactly $0$, since this is the defining relation for $\beta_h$. For $n\geq 2$, owing to behaviour of $a_n$ at large central charge \eqref{a_n-behave}, we have
\bea\label{compare2}
\bigg(\langle\psi| \mathbbm{B}_{n}|\psi\rangle_{\text{cylinder}}-\langle \mathbbm{B}_{n} \rangle_{\text{thermal}}\bigg) \ \underset{c\rightarrow\infty}{\underset{L \rightarrow \infty}{=}} \ \left(\frac{h}{L^{2}}\right)^{n}\mathrm{O}(1/c).
\eea

These operators which violate global ETH \eqref{thermalization} for finite $c$, are related to energy conservation.
In particular the violation is due to the $\mathrm{O}(c^{-1})$ deviations of $ \langle \mathbbm{B}_{n}\rangle_{\text{thermal}}$ from $\langle \mathbbm{B}_{1} \rangle^{n}_{\text{thermal}}$. In thermodynamic limit as $L\rightarrow\infty$ and $h\sim L^{2}$, we have $ \langle\psi| \mathbbm{B}_{n}|\psi\rangle_{\text{cylinder}}-\langle \psi|\mathbbm{B}_{1}|\psi \rangle^{n}_{\text{cylinder}} =0$. If we define, 
$$
\sigma(A_n)_{\alpha} \ \equiv \  \Tr (  A_n \rho_\alpha ) -  \bigg(\Tr ( A_1 \rho_\alpha )\bigg)^n,
$$
then for the operators, $\mathbbm{B}_n$ we have,
\beq
\sigma(\mathbbm{B}_n)_{\psi} - \sigma(\mathbbm{B}_1)_{\beta }  \underset{c\rightarrow\infty}{\underset{L \rightarrow \infty}{=}} \ \left(\frac{h}{L^{2}}\right)^{n}\mathrm{O}(1/c).
\eeq
For $n=2$ the above reduces to a version of the energy density variance considered in \cite{Garrison:2015lva}. In general for $n\geq2$, $\underset{z \in \mathcal{A}}{\int} dz \mathbbm{B}_n(z)$ plays the analogue role of $H_\mathcal{A}^n$ in conjecture \ref{conj2}.

The identity module also contains other quasi-primaries, $\bt_{k/2}^{(j)}$ at positive integer levels $k$, not containing $\normord{T^{\lfloor k/2 \rfloor}}$\footnote{There can be more than one quasi-primary at level $k$, indexed by $j$. Quasi-primaries of this kind appear from level 6 onwards. The generating function for the number of quasi-primaries at each level is 
 \begin{align}
 \chi_{\text{quasi}} &= (q^{c/24}\chi_{\text{vac}}-1)(1-q) = q^2 +q^4 + 2q^6 + 3q^8+q^9 + 4q^{10} + \cdots \nonumber
 \end{align}
  here $\chi_{\text{vac}}$ is the vacuum character given by $q^{-c/24}\prod_{n=2}^\infty (1-q^n)^{-1}$.}. The expectation value $\langle \psi | \bt_{k/2}^{(j)} | \psi \rangle_{\text{cylinder}}$ vanishes in large $L$ limit since it goes like $h^l/L^k$ with  $l <\lfloor k/2\rfloor$. Moreover, in that scenario, $a_{ k/2}^{(j)}$ goes as $c^{p}$ with $p<\lfloor k/2\rfloor-1$. The origin of the $-1$ can be understood as follows. The operator $\bt_{k/2}^{(j)}$ contains $p$ stress tensors, with $p<\lfloor k/2\rfloor$ and derivatives to make up the dimensions. The derivatives acting on $c^p s^p$ gives zero as $s$ is a constant. It can then be easily seen that the leading term in the large $c$ limit will be given by, $c^{p-1}$. This implies that, $\langle \psi | \bt_{k/2}^{(j)} | \psi \rangle_{\text{cylinder}} - \vev{\bt_{k/2}^{(j)} }_{\text{thermal}}$ goes to $0$ in large $c$ limit as $1/c^m$ with $m\geq 2$. 

Therefore we have shown that all the quasi-primaries in the identity block satisfy global ETH in the large central charge limit. Note that there is a specific order in which limits are being taken here : the large $h$ approximation is used first and then that of large $c$. 

Global ETH is also violated by  non-vacuum primaries (and hence their descendants) for finite $L$. However, as we shall see later, in the thermodynamic limit, $L \rightarrow \infty$, these are typically exponentially suppressed unlike that for $\mathbbm{B}^\nj_{n\geq2}, \bt_{k/2\geq3}^{(j)}$. 

\def\cyl{{\mathbb{R}\times \mathbb{S}^1_L}}
\def\torus{{\mathbb{T}^2}}

\section{Trace square distance in the short interval expansion }\label{section:tracedistance}

\subsection*{Setting up the short interval expansion}
As a diagnostic of subsystem ETH, we shall consider the square of the trace two norm distance, 
\beq
\T = ||\rho_A^\psi - \rho_\cA^{\beta}||_2^2  = \Tr (\rho^\psi_\cA)^2 + \Tr (\rho^{\beta}_\cA)^2 - 2~\Tr \rho_\cA^{\beta} \rho_\cA^\psi. \label{all1}
\eeq
In the path integral formalism, each of the terms appearing above  admits a representation in terms of partition functions of the CFT of two Riemann surfaces glued along the cut defined by the subsystem $\cA$.
This is depicted in Fig.~\ref{fig:tracedis}. 
 For the first two terms, 
the two surfaces are identical. The first term is a path-integral in a spatial cylinder ($\cyl$) joined to another copy, whilst the second one is a torus ($\torus$ with modular parameter $\tau=i\beta/L$) joined to another copy. The last term, which is cross-term involving $\rho_\cA^\beta$ and $\rho_\cA^\psi$, corresponds to the glued cylinder and torus. If we denote the partition function of two glued Riemann surfaces $\Sigma_1$ and $\Sigma_2$ as $Z_{\Sigma_1,\Sigma_2}$, we have
\begin{align}\label{glued-Riemann}
\T = Z_{\cyl,\cyl} + Z_{\torus,\torus} - 2 Z_{\cyl,\torus}.
\end{align}
Since we are interested in reduced density matrix corresponding to energy eigenstate $\ket{\psi}$, there are operator insertions on the cylinders. In the high temperature regime, the manifold corresponding to the third term becomes a spatial cylinder joined to a thermal cylinder \ie a cylinder with compactified Euclidean time. This is the approximation $\torus \cong \mathbb{R}\times \mathbb{S}^1_\beta$, when finite-size corrections (arising from the $q$-expansion) are ignored. 

 \begin{figure}[t]
 	\begin{overpic}[scale=0.35]{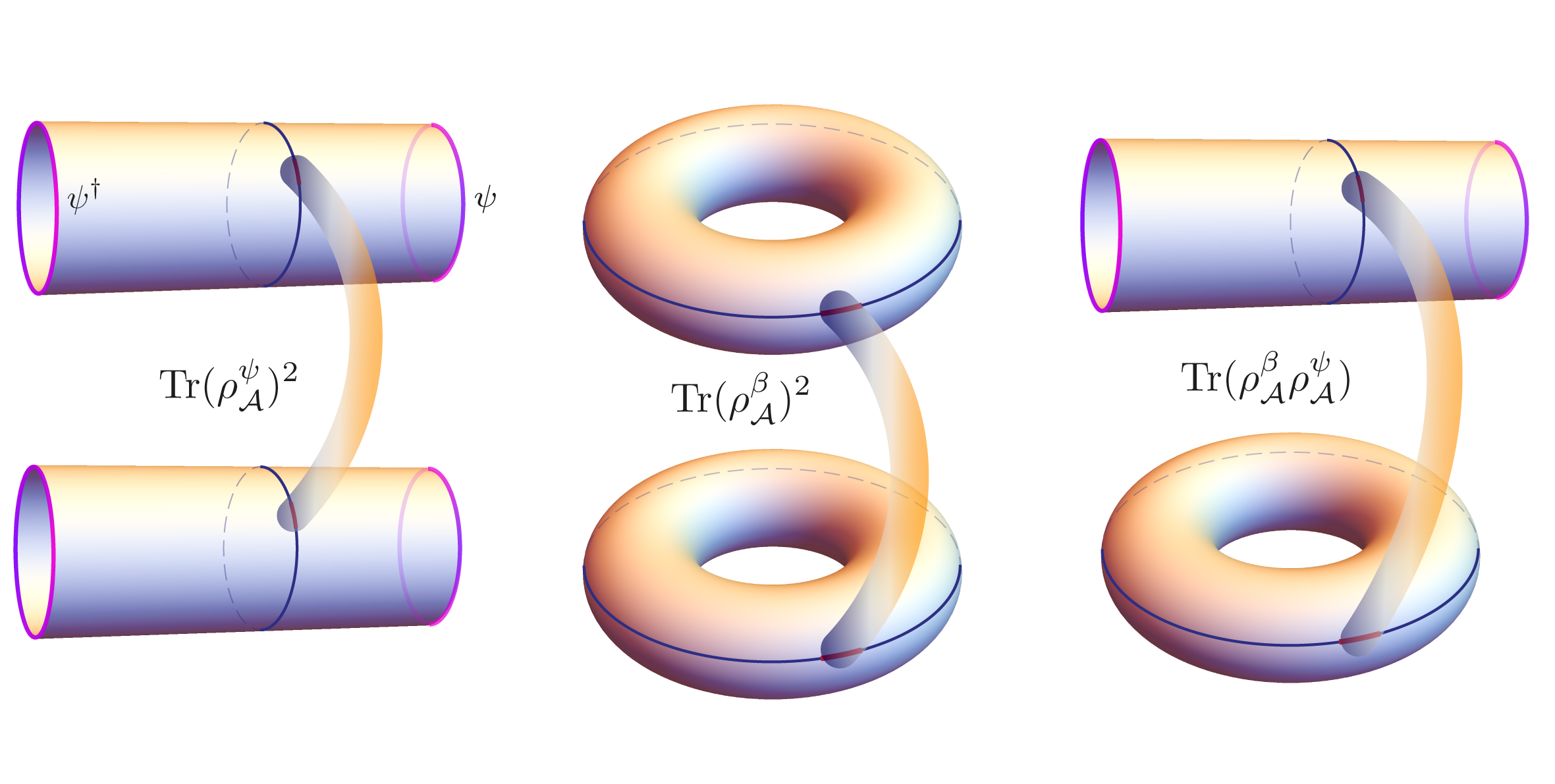}
 	\end{overpic}
 	\centering
 	\caption{Sewed Riemann surfaces corresponding to the path integral representation of the terms in the trace square distance, equation \eqref{all1} and \eqref{glued-Riemann}.} 
 	\label{fig:tracedis}
 \end{figure}

The standard way to compute correlators on multiply connected Riemann surfaces such as the terms appearing in \eqref{all1} is via correlation functions of  twist fields \cite{Cardy:2007mb, Calabrese:2009qy}. This method is straightforward when each of the Riemann sheets correspond to genus-0 surfaces. Such terms were calculated using the twist correlator in \cite{Sarosi:2016oks} who computed $\T$ between primary eigenstates $\psi, \phi$. The cross-terms of the kind in \eqref{glued-Riemann} also appeared in the context of studying thermalization in critical quantum quenches an infinite strip attached to a torus \cite{Cardy:2014rqa, Mandal:2015jla}. This is calculated using the method of short interval expansion in the high temperature regime which we shall also deploy here. This technique was introduced in the context of Renyi entropies of disjoint intervals in 2d CFTs in \cite{Headrick:2010zt}. 

The short interval expansion is an expansion in powers of ratio of subsystem size to any other relevant length scale of the CFT, for example total system size $L$ and/or inverse temperature $\beta$.  
If the size of the subsystem is small enough, this allows us to implement \textit{cutting and sewing} of the original path integral over the two sheeted manifold  in terms of path integral over one sheeted manifold \cite{Sen:1990bt,Gaberdiel:2010jf,Headrick:2015gba} (see in particular \S(3.6) of \cite{Sen:1990bt} and references therein).

The cuts, defined by disks surrounding the subsystem $\cA$, creates a complete set of orthogonal states on each of the sheets. The individual sheets are then sewn along the cuts. This is performed by inserting a complete set of states, which match the boundary conditions at the two cuts.  This amounts to inserting
\begin{align}
\ket{\mathcal{I}} = \sum_{k_1,k_2} C_{k_1,k_2} \ket{\phi_{k_1}} \otimes \ket{\phi_{k_2}},
\end{align}
where, the complete set, $\lbrace \phi_k\rbrace$, consists of the quasi-primaries of the CFT\footnote{{This may seem like a sneaky assumption. However, the contribution from the non-quasi-primaries can be shown to be sub-dominant in the thermodynamic limit. For example, $\vev{\partial T}_\beta=0$ due to the transformation rules of $\partial T$ and $\vev{\partial T}_\psi\sim h/L^3\underset{h\sim L^2}{\underset{L \rightarrow \infty}{\rightarrow}}0$.}}. Therefore, the  short interval expansion is effectively a sum over the the complete set of quasi-primaries. The coefficients $C_{k_1,k_2}$ are given by
\begin{align}\label{ck1ck2-def}
C_{k_1,k_2} = \frac{Z_2}{(Z_1)^2} \lim\limits_{z_{j}\rightarrow \infty_j} z^{2(\Delta_1+\Delta_2)} \bar{z}^{2(\Delta_1+\Delta_2)} \vev{\phi_{k_1}(z_1)\phi_{k_2}(z_2)}_{\Sigma_{g_1,g_2}}.
\end{align}
Here, $Z_n$ is the partition function of the $n$-sheeted plane. The coefficients are symmetric under $k_1$ and $k_2$ exchange. We provide further details in  Appendix \ref{app-c}.  The partition functions of interest, equation \eqref{glued-Riemann}, can then be rewritten as 
\begin{align}
Z_{\cyl,\cyl} &=  \sum\limits_{k_1,k_2} C_{k_1 k_2} \vev{ \psi \phi_{k_1} \psi^\dagger}_{\cyl} \vev{ \psi \phi_{k_2} \psi^\dagger}_{\cyl}. \nn
Z_{\torus,\torus} &=  \sum\limits_{k_1,k_2} C_{k_1 k_2} \vev{   \phi_{k_1}  }_{\torus}\vev{   \phi_{k_1}  }_{\torus}.  \\
Z_{\cyl,\torus} &=  \sum\limits_{k_1,k_2} C_{k_1 k_2} \vev{ \psi \phi_{k_1} \psi^\dagger}_{\cyl}\vev{   \phi_{k_1}  }_{\torus}. \nonumber
\end{align} 
The short interval expansion for $Z_{\cyl,\torus}$ is depicted in Figure~\ref{ins}.

\begin{figure}[t]
\centering
\begin{overpic}[scale=0.4]{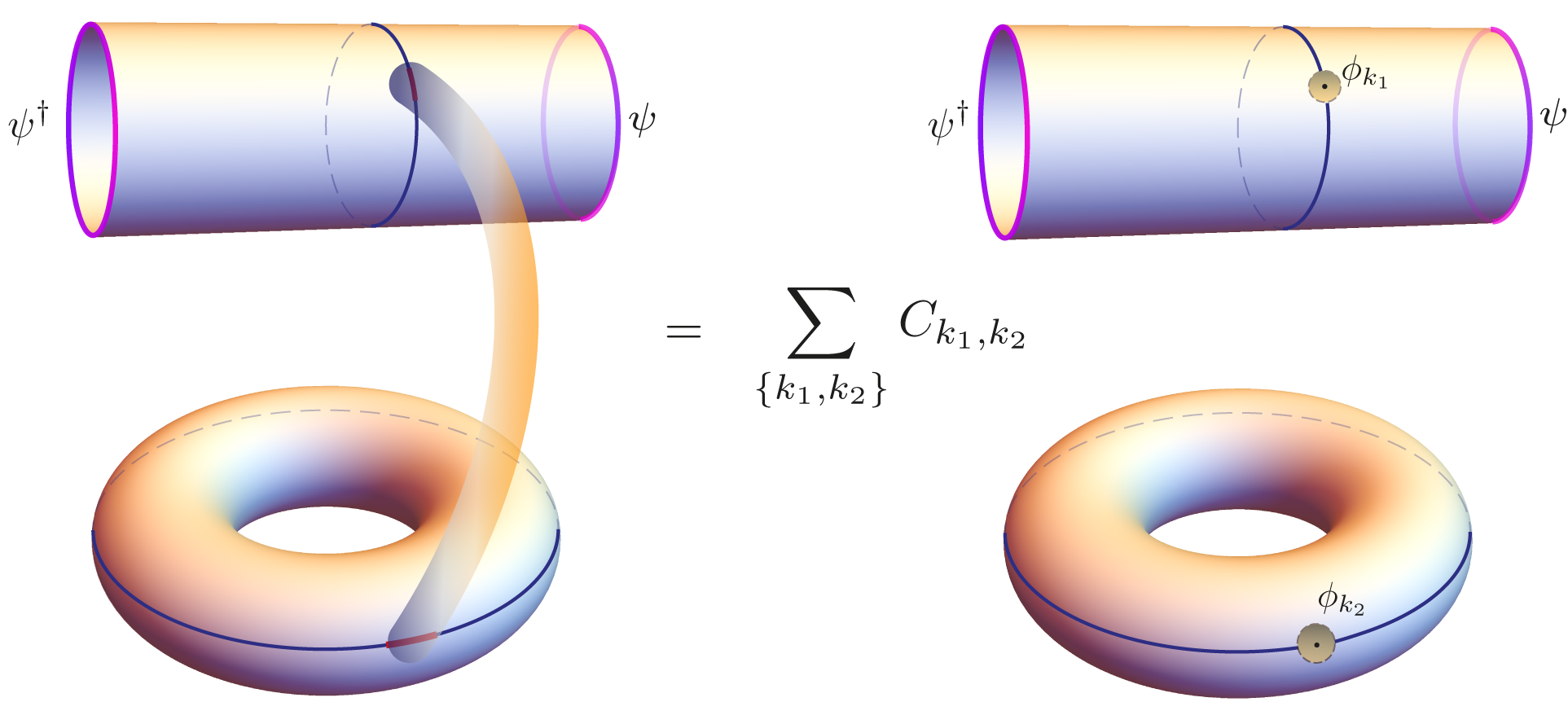}
\end{overpic}
\caption{Short interval expansion of $Z_{\cyl,\torus}$ : A small disk of radius of $\ell$ (corresponding to the subsystem $\cA$) is cut, subsequently the sewing is performed by inserting a complete set of states i.e $\ket{\mathcal{I}}$} \label{ins}
\end{figure}

The trace square distance now takes the following form 
\begin{align}\label{simpler}
\T = \sum\limits_{k_1, k_2\setminus 0} C_{k_1 k_2} \bigg( \vev{ \psi \phi_{k_1} \psi^\dagger}_{\cyl}  -\vev{\phi_{k_1} }_{\torus} \bigg) \bigg( \vev{ \psi \phi_{k_2} \psi^\dagger}_{\cyl}  -\vev{\phi_{k_2} }_{\torus} \bigg).
\end{align}
The coefficients $C_{k_1,k_2}$, \eqref{ck1ck2-def}, has the right powers of the subsystem size, $\ell$ which make each term in the sum dimensionless. The location of $\phi_k$'s are at the centre of the interval $[0,l]$ in each geometry \ie at  $(w,\bar{w})_\cyl=(\frac{\ell}{2},\frac{\ell}{2})$. 

Note that the trace in the two norm distance \eqref{simpler} contains information about the deviations from global ETH. The corrections to  \eqref{thermalization} appears naturally for the local operators $\phi_{k_1}$ and $\phi_{k_2}$. Defining the  \textit{deviation} for a general operator $\mathbbm{O}$ as
\begin{align}\label{deviation-def}
A_{\mathbbm{O}} = \vev{\psi^\dagger \, \mathbbm{O} \,\psi }_\cyl - \la\mathbbm{O}\ra_\torus.
\end{align}
we can rewrite the trace square distance as
\begin{align}\label{simpler2}
\T = \sum\limits_{k_1, k_2\setminus 0} C_{k_1 k_2} A_{\phi_{k_1}} A_{\phi_{k_2}}.
\end{align}

\subsection*{Limits and the regime of validity of our analysis}
As mentioned earlier, the short interval expansion is a good approximation when the extension of size of the interval, $\ell$ is smaller than all other length scales, ($L$ and $\beta$). 
Furthermore, as noted earlier to have finite energy density ($E\sim h/L^2$) eigenstates we require $h\gg 1$. This implies, $\beta/L\sim\sqrt{c/24h}$ is small. This is the high temperature regime with $l \ll \beta \ll L$. In this regime the leading contribution to the torus one point function comes from the cylinder (of circumference $\beta$) and has exponentially suppressed finite-size corrections. In terms of dimensionless ratios, $r = \ell/L, s = L/\beta$, we are calculating perturbatively in the \textit{subsystem fraction}, $r$. The finite-size corrections to our calculation goes like $e^{-s}$.  

In what follows, we shall keep the central charge fixed, and our results are valid for any value of $c$. However, we shall also investigate the `holographic' regime, where $c$ is large. 

\subsection*{Evaluating $\T$}

At this point we may separate out the contribution of the quasi-primaries corresponding to the identity block which is universal for all CFTs. Ignoring the exponentially suppressed finite-size corrections the first few deviations $A_{\mathbbm{O}}$ read, (see \S\ref{app-b} for details),
\bea
A_T &=&  \frac{\pi^2 c}{6 L^2} + \frac{\pi^2 c}{6 \beta^2} - \frac{ 4 \pi^2 h}{L^2}, \label{AT} \\ 
A_\Omega &=&  \bigg(  \frac{\pi^{2}c}{6 L^2} - \frac{ 4 \pi^2 h}{L^2} \bigg)^2  - \frac{ \pi^{4}c^2}{36 \beta^4}, \label{Ak}\\
A_\Lambda &=& -\frac{22c  + 5c^2}{180\beta^4}\pi^4 
+ \frac{22c  + 5c^2}{180L^4}\pi^4 + \frac{ 16 \pi^4 h^2 }{L^4} - \frac{  h \pi^4}{3 L^4}( 80 + 4c ). \label{AL}
\eea
where $T$ is the holomorphic part of the stress-tensor and  $\Omega = \normord{T \bar{T}}$, $\Lambda = \normord{T T} - \frac{3}{10} \partial^2 T$ are the quasi-primaries at level 4\footnote{By $\normord{\Theta \Theta'}$ we mean that divergences arising due to contact-terms are subtracted out.}. For the anti-holomorphic pieces, the expressions for  $A_{\bar{T}}$ and $A_{\bar{\Lambda}}$ stay the same with $h$ replaced by $\bar{h}$.  Some of the coefficients are derived and listed in \S\ref{app-c}. 

Grouping the contributions of the trace square distance $\T$, given in  \eqref{simpler2}, from the vacuum module, we have an expansion in the subsystem fraction, $r(=\ell/L)$
\bea\label{id}
\T^{(\text{vac})} = &\null& \hspace{-.4cm}C_{T,T} A_T^2 + 2 C_{T,\bar{T} } A_T A_{\bar T} + C_{\bar{T} ,\bar{T}} A_{\bar T}^2 
+   2 C_{T, \Omega} A_T A_\Omega + 2 C_{T, \Lambda} A_T A_\Lambda  \nn
 &+& 2 C_{T , \bar\Lambda} A_T A_{\bar\Lambda} + 2 C_{\bar T, \Omega} A_{\bar{T}} A_\Omega + 2 C_{\bar T, \Lambda} A_{\bar{T}} A_\Lambda + 2 C_{\bar{T} , \bar\Lambda} A_{\bar{T}} A_{\bar\Lambda}  + C_{\Lambda , \Lambda} A_{\Lambda}^2 \nn
 &+& C_{\bar \Lambda, \bar \Lambda}A_{\bar{\Lambda}}^2 +2 C_{\Omega, \Lambda} A_\Omega A_\Lambda + 2 C_{\Omega, \bar\Lambda} A_\Omega A_{\bar\Lambda} +  C_{\Omega, \Omega} A_\Omega^2 + 2 C_{\Lambda, \bar\Lambda} A_\Lambda A_{\bar\Lambda} \\
 &+& 2 C_{\mathcal{B},T} A_T A_{\mathcal{B}} + 2 C_{\bar{\mathcal{B}},\bar{T}} A_{\bar{ T}} A_{\bar{ \mathcal{B}}} +2 C_{\mathcal{D},T} A_T A_{\mathcal{D}} + 2 C_{\bar{ \mathcal{D}},\bar T} A_{\bar T} A_{\bar{ \mathcal{D}}} \nn
 &+& 2 C_{\bar{ \mathcal{B}},T} A_T A_{\bar{ \mathcal{B}}} + 2 C_{ \mathcal{B},\bar T} A_{\bar T} A_{ \mathcal{B}} +
 2 C_{\bar{ \mathcal{D}},T} A_T A_{\bar{ \mathcal{D}}} + 2 C_{ \mathcal{D},\bar T} A_{\bar T} A_{ \mathcal{D}}+  \dots \nnn
\eea
Here, $\mathcal{B} = \normord{\partial T \partial T} - \frac{4}{5} \normord{ \partial^2 T T } - \frac{1}{42} \partial^4 T$ and $\mathcal{D} = \normord{ T\normord{TT}} - \frac{9}{10} \normord{\partial^2 T T} -\frac{1}{28} \partial^4 T + \frac{93}{70c + 29} \mathcal{B}$ are holomorphic quasi-primaries at level 6. The first three terms and the next six in \eqref{id} go as $r^4$ and $r^6$ respectively. The last fourteen terms behave as $r^8$. The $\dots$ are terms with higher even powers of $r$ and are thus suppressed in $r\rightarrow 0$ limit. 
Using \eqref{AT}, \eqref{Ak}, \eqref{AL} and \eqref{Ckj} we can write down the explicit form of the universal answer.  Note that in the case of the spinless eigenstate the $r^4$ and $r^6$ terms are proportional to $A_T$. 

The theory dependent contributions to the trace square distance come from the set of primaries ($\phi_{\Delta,s}$ ) and their descendants which are of the form  
$$\CO_{\Delta,s,N,\bar{M}} = L_{-n_1} L_{-n_2} \dots L_{-n_\alpha} \bar{L}_{-m_1} \bar{L}_{-m_2} \dots  \bar{L}_{-m_\gamma} \phi_{\Delta,s},$$ 
with $\sum_{i=1}^\alpha n_i = N$ and $\sum_{j=1}^\gamma m_j = M$. 
The coefficient  $C_{\CO_1, \CO_2}$ and the deviation $A_\CO$  in the term $C_{\CO_1, \CO_2} A_{ \CO_1}  A_{\CO_2}$ contributing to $\T$ has the behaviours (ignoring finite-size corrections)
$$C_{\CO_1,\CO_2}\sim \ell^{2 (\Delta_{\CO_1}+N_1 + \bar{M}_1)} \delta_{\phi_1\phi_2} \delta_{N_1 N_2}\delta_{M_1 M_2}, \qquad A_\CO \sim L^{-2(\Delta_\CO + N + M)}.$$ 
 This shows that the non-universal piece is also a perturbative series in $r$, with a general term behaving like $r^{2(\Delta_\CO + N + M)}$. Explicitly for the primary $\phi$ with conformal dimension $\Delta_\phi$ and spin $s_\phi$ we have, 
\bea  \label{Aphi}
A_\phi &=& c_{\psi \phi \psi} \bigg( \frac{2\pi}{L} \bigg)^{\Delta_\phi} e^{-\frac{ i \pi}{2} s_\phi }. 
\eea
All descendants, $\CO_{\Delta_\phi, s_\phi, N_\phi, M_\phi}$ of $\phi$ will also have the three point coefficient $c_{\psi \phi \psi}$ in the corresponding $A_{\CO}$.  

\section{Deviations from subsystem ETH}\label{section:deviationfromETH}

In this section, we examine the deviation of the trace square distance from zero for spin-less finite energy density  eigenstates. To begin with, we shall ignore finite size effects. Subsequently, we shall consider them and show how it affects the trace distance.
\subsection{The thermodynamic limit}
 The thermal density matrix, that seems to approximate the spinless eigenstate with the finite energy density, is at inverse temperature $\beta_h$ \eqref{beta1}.  As can be already seen from the structure of the universal terms \eqref{id}, this forces the coefficient of the $r^4, r^6$ terms to zero. At $\beta = \beta_h$ the leading non-universal terms also have a specific decay. 

\subsubsection*{Universal part}
For the universal contribution \eqref{id} at $\beta = \beta_h$, the leading contribution comes from the $r^8$ term,
\beq
\hat \T^{(\beta = \beta_h)}_{\text{univ}}  = \frac{ \T^{(\beta = \beta_h)}_{\text{univ}} }{S_{2,A}^{vac} }= 
\pi^8 r^8 \frac{h^2(5 c^2+42c+108)(13 c+44 h)^2}{25600 c^3 (5 c+22)} + \CO(r^{10}).
\eeq
we have normalized the square of the trace two norm by the second Renyi entropy of the CFT vacuum on the plane denoted by $S_{2,A}^{vac} = Z_2/ Z_1^2 = c_2 (\ell/\epsilon)^{-c/4}$, where $\epsilon$ is the UV cut-off. This answer is valid for all values of $c$. Note that in the thermodynamic (large $L$) limit, only the $h^{4}$ term survives and the trace distance goes as 
$$\hat \T^{(\beta = \beta_h)}_{\text{univ}}  \simeq  \frac{121(5 c^2+42c+108)}{1600 c^3 (5 c+22)} (h\pi^{2}r^{2})^{4}.$$
 In large $c$ limit, this goes to zero as $$\hat \T^{(\beta = \beta_h)}_{\text{univ}}  \simeq  \frac{121}{1600 c^2} (h\pi^{2}r^{2})^{4}.$$
The contributions to the trace distance coming from all the quasi-primaries of identity block goes to $0$ in large $c$ limit, as evident from the argument presented in subsection~\S\ref{ETH violating operators}. \\

\subsubsection*{Non-universal part}
When the CFT has light operators $\phi$ with $\Delta_\phi < 4$, the leading behavior at $\beta=\beta_h$ comes from the non-universal sector which goes like $r^{2 \Delta_\phi}$. Recall from \eqref{Aphi} that the leading non-universal pieces are proportional to the heavy-light-heavy three point coefficients, $c_{\psi \phi \psi}$. Recently using modular covariance properties of torus one point functions, \cite{Kraus:2016nwo} estimated the typical three point coefficients of the type, $c_{\psi \phi \psi}$ resulting in a Cardy like formula valid for large $h$  \footnote{In this subsection, by `typical' we mean the light primaries which have the structure constant close to the mean value calculated in \cite{Kraus:2016nwo}.}. The estimate is in terms of $c_{\chi \psi \chi}, \Delta_\chi = h_\chi + \bar{h}_\chi $ where $\chi$ is the lightest operator with non-vanishing three-point coefficient. The form of the average heavy-light-heavy coefficient is constrained to be of the type,
\beq\label{km}
{c_{\psi \phi \psi }}_{av}   = \frac{c_{\chi \phi \chi }}{\rho(h) } T_\phi(h) .
\eeq
where, 
\beq \label{Th}
T_\phi(h) \approx  N_\phi  \bigg( 2h - \frac{c}{12} \bigg)^{\frac{\Delta_\phi}{2} -\frac{3}{4}} \exp \bigg\{ 4\pi \sqrt{ \bigg( \frac{c}{12} - \Delta_\chi \bigg) \bigg( 2h - \frac{c}{12} \bigg) + \dots}\bigg\}.
\eeq
Here $N_\phi$ is independent of $h$ and the $\dots$ in exponent go to zero as $h \rightarrow \infty$. Now we multiply this with the inverse of the density of states ($\rho(h)^{-1} \sim e^{-S(h)}$) appearing in \eqref{km}. At this point we may use the Cardy formula\footnote{ We use the Cardy formula along with the prefactor coming from integrations around saddle, $$\rho(h ) \approx  \bigg( 2h - \frac{c}{12} \bigg)^{-3/4} \exp\bigg\{ 4\pi \sqrt{ \frac{c}{6} \bigg( h - \frac{c}{24} \bigg) } \bigg\}.$$} for the density of states to simplify the above formula further to, 
\beq\label{cav}
 {c_{\psi \phi \psi }}_{av}  =  c_{\chi \phi \chi } N_\phi  \exp \bigg\{ -4\pi \sqrt{\frac{c}{6}\bigg( h - \frac{c}{24} \bigg) } \bigg( 1 - \sqrt{1 - \frac{12 \Delta_\chi}{c}} \bigg) + \frac{\Delta_\phi}{2}\log\bigg( 2h - \frac{c}{12} \bigg)\bigg\}.
 \eeq
 We see that in the  $h \gg c \gg \Delta_\chi$ limit, the leading non-universal part of the squared TSD is,
 \beq\label{non-uniT}
 \hat \T^{(\beta= \beta_h)}_{\text{non-univ}} = \frac{\T^{(\beta= \beta_h)}_{\text{non-univ}} }{S_{2,A}^{vac} } \sim r^{2\Delta_\phi} c_{\chi \phi \chi}^2 \exp\bigg( -8 \pi \Delta_\chi \sqrt{ \frac{ 6h}{c} } \bigg)(2h)^{\Delta_\phi/2} .
 \eeq
 A caveat here is that since the three point coefficients are not positive definite, while the average heavy-light-heavy coefficients decay for $h \rightarrow \infty$, there may be some $c_{\psi \phi \psi}$ which do not decay. 

  It is interesting to note that when $\Delta_\chi = c/12$, in the $h\rightarrow \infty$ limit the non-universal contribution goes like $\sim r^{2\Delta_\phi} c_{\chi \phi \chi}^2 e^{-4\pi \sqrt{c/6(h - c/24)}}$. Using the Cardy formula for the density of states, we may identify this exponential entropic suppression as $e^{-\mathcal{S}}$.  
\subsection{Finite size effects }
Throughout the above analysis we have ignored the finite size corrections which are present in the torus one point functions. However conformal symmetry constrains the finite size corrections to take a specific form. For instance for the finite size effects in the torus one point function for the stress tensor is given by \eqref{Ttorus}. If we take the leading correction into account then the equation that one needs to solve for $\beta_h$ gets modified to, 
\beq
\frac{ L^2}{\beta^2}  \bigg( c - 48 e^{-L/\beta} \bigg) = 24h - c.
\eeq
 We can solve this perturbatively around $\beta_{h,0} = \frac{L}{\sqrt{ 24h/c-1}}$ in the small parameter, $e^{-L/\beta_{h,0}}$. This gives the corrected value of $\beta_h$, 
 \beq\label{beta11}
 \beta_{h,1}  = \frac{L}{\sqrt{ \frac{24h}{c} -1} } \sqrt{ 1 - \frac{48}{c} \exp\bigg( - 4\pi c \sqrt{ \frac{24h}{c} -1} \bigg) }.
 \eeq
Note that the correction to $\beta_h$ is suppressed by $\frac{1}{c} e^{-8\pi\sqrt{6 h c } }$ in the limit $h \gg c \gg 1 $. In the universal contribution to the trace distance squared, the leading term will again be $\propto r^8$ but will contain corrections suppressed by powers of $e^{-\sqrt{hc} }$.

In the non-universal contribution to $\T$, the one point function of a primary $\mathbbm{O}$ on the torus also admits a $e^{-L/\beta}$ expansion and is proportional to the coefficient $c_{\chi \CO \chi}$ where $\chi$ is the lowest primary with non-zero coefficient. 
\beq\label{Oth}
\frac{\vev{ \opr}_{\torus} }{\Tr (e^{-\beta H}) }  \approx c_{\chi \opr \chi} \bigg( \frac{ 2\pi}{\beta} \bigg)^{\Delta_{\opr} } e^{-2\pi \Delta_\chi  L/\beta  } + \dots.
\eeq 
The $\dots$ are contributions from other light primaries $\eta$ with $\Delta_\eta > \Delta_\chi$ and hence are suppressed in the $\beta_h \rightarrow 0 $ limit. The $\beta_{h}$ has finite size correction as given in \eqref{beta11}. Since the corrections in $\vev{\opr}_{\torus}$ are already suppressed exponentially for small $\beta$ in \eqref{Oth}, we can neglect the finite size corrections to $\beta_h$ itself.  To proceed further, we assume that there is a gap above $\chi$. Thus including leading finite size corrections, the non-universal part of the squared TSD goes as, 
\bea
\hat \T^{(\beta= \beta_h)}_{\text{non-univ}} = \frac{\T^{(\beta= \beta_h)}_{\text{non-univ}} }{S_{2,A}^{vac} } &\sim& 
 r^{2\Delta_\phi} \bigg( c_{\psi \phi \psi} - c_{\chi \phi \chi} e^{-2\pi s_h \Delta_\chi +  \Delta_\chi \log s_h }\bigg)^2, \\
 &\sim& r^{2\Delta_\phi} c_{\chi \phi \chi }^2 \bigg( \frac{ T_\phi(h) }{\rho(h) } - e^{  -2\pi \Delta_\chi  \sqrt{24h/c-1} + \frac{\Delta_\chi }{2} \log ( 24h/c -1 ) }\bigg)^2.\nnn
 \eea
 where in the last line above we approximated the heavy-light-heavy coefficient by its average value \eqref{cav}. Once again $\phi$ is the lightest primary in the spectrum that has a non vanishing three point coefficient $c_{\psi \phi \psi}$. $\chi$ is the lightest primary with non vanishing $c_{\chi \phi \chi}$.  In terms of $h$ and $c$ this expression is,
 \bea
\hat \T^{(\beta= \beta_h)}_{\text{non-univ}}&\sim& r^{2\Delta_\phi} c_{\chi \phi \chi}^2 \bigg[  \exp \bigg\{ -4\pi \sqrt{\tfrac{c}{12}\bigg( 2h - \tfrac{c}{12} \bigg) } \bigg( 1 - \sqrt{1 - \tfrac{12 \Delta_\chi}{c}} \bigg) + \tfrac{\Delta_\phi}{2}\log\bigg( 2h - \tfrac{c}{12} \bigg)\bigg\} \nn
&\null& \qquad\qquad \quad   -  \exp\bigg\{    -2\pi \Delta_\chi  \sqrt{24h/c-1} + \frac{\Delta_\chi }{2} \log ( 24h/c -1 ) \bigg\}  \bigg]^2.
 \eea
 In the $h\gg c\gg \Delta_\chi$ limit, the behavior is,
 \beq
\hat \T^{(\beta= \beta_h)}_{\text{non-univ}}\sim  r^{2\Delta_\phi} c_{\chi \phi \chi}^2  \exp\bigg( -8\pi \Delta_\chi \sqrt{ \frac{ 6h}{c} }\bigg) \bigg( (2h)^{\Delta_{\phi/2}} - (24h/c)^{\Delta_\chi/2} \bigg)^2.
 \eeq
 Thus in this limit, the qualitative behavior \eqref{non-uniT} remains unchanged  even the finite size corrections are included.

\def\cV{\mathcal{V}}

\def\cV{\mathcal{V}}

\section{Conclusions}\label{sec:last}

In this work, we have tested eigenstate thermalization in the context of conformal field theories in two dimensions. In particular, we have focused on the sector which is universal to all 2d CFTs, \ie the vacuum Virasoro module. For the class of eigenstates, which we have considered, our analysis yields that the quasiprimaries satisfy global ETH strictly in the large central charge limit, the deviations being suppressed in powers of $1/c$.

We have also quantified the deviations from subsystem ETH by calculating the trace distance squared between the energy eigenstate and the related hypothesized thermal state, in the short interval expansion. The universal deviations appear at the order $r^8$ in the short interval expansion. Furthermore, these deviations are suppressed for large central charge. The leading non-universal deviations are dictated by the conformal dimension of the lightest operator in the CFT spectrum. 

In hindsight, the deviation of ETH away from $c\rightarrow\infty$ is not too surprising. The central charge serves a measure of the number of degrees of freedom which is indeed required to be large for eigenstate thermalization.  Furthermore, as is well known,  rational CFTs (typically at small values of central charge) are rendered integrable, owing to the KdV hierarchy present in each Verma module \cite{Bazhanov:1994ft, Sasaki:1987mm, deBoer:2016bov}. Hence, we do not expect ETH to hold good for RCFTs. Moreover, there is no known generalization of such integrable structure to be present at generic values of central charge and/or for irrational CFTs. In particular, pinning down the structure constants and spectra of generic CFTs is a difficult task and integrability in the sense of \cite{Bazhanov:1994ft} does not help.\footnote{We thank Christoph Keller and Tom Hartman for clarifications on this.} 
One can note, however, that if additional conserved quantities are present in the theory due to an enhancement of the chiral algebra, a suitably modified version of ETH formulated in terms of the generalized Gibbs ensemble is expected to hold true \cite{Mandal:2015jla, Cardy:2015xaa, Sotiriadis:2015kfa}.

These results at large central charge (once combined with some assumptions on the CFT spectrum) potentially provide sharp implications for holographic CFTs. Our findings show that that the finite energy density state mimics the thermal state (dual to the BTZ black hole in the bulk) to a good approximation, if $1/c$ corrections and finite-size effects are ignored. Furthermore, a holographic route to calculate the trace square distance would be to consider handlebody geometries (dual to glued Riemann surfaces in Fig.~ \ref{fig:tracedis}), which are orbifolds of $AdS_3$ by the Schottky group. This correspondence has been used to reproduce the Renyi entropies (including finite-size corrections)  in \cite{Faulkner:2013yia,Barrella:2013wja,Datta:2013hba}. 
 Assuming $\psi$ to be a heavy excited state, the gravity dual is given by a conical defect, wherein the mass of the conical defect is related to the conformal dimension of the heavy operator as 
$
 m = 2 \sqrt{h(h-1)} \cong 2h \gg 1 \nonumber
$.
  On the other hand, the thermal state is holographically described  the BTZ black hole. Denoting $Z_{i,j}$ as the handlebody geometry which glues geometry $i$ and geometry $j$ along the subsystem, we are, therefore, led to the following schematic relations
  \begin{align*}
  \Tr ({\rho^\psi_\cA})^2  = Z_{\text{conical,conical}} \qquad 
  \Tr ({\rho^{\beta}_\cA})^2  = Z_{\text{BTZ,BTZ}} \qquad
  \Tr (\rho_A^{\beta} \rho_\cA^\psi) = Z_{\text{BTZ,conical}}
 \end{align*}
 In addition to evaluating the classical Zograf-Takhtajan action for the handlebody geometries \cite{Krasnov:2000zq,Faulkner:2013yia}, it would also be of interest to evaluate the one-loop corrections to the same \cite{Barrella:2013wja}. Note that in the thermodynamic limit, the global ETH deviations responsible for the leading universal subsystem ETH deviations are of the form, $\vev{ \normord{T^n} }_{\text{BTZ}}  - \vev{ T}_{\text{BTZ}}^n$. From the bulk this deviation can be obtained from correlators using the GKPW prescription in the BTZ black hole and the conical defect \cite{Gubser:1998bc, Witten:1998qj}. Alternatively, one could generalize the analysis of \cite{Bagchi:2015wna} to the dual backgrounds to obtain the same. 

It would be of potential interest to investigate the subsystem ETH in higher dimensional CFTs. One can hope to generalize the Cardy formula for structure constants \cite{Kraus:2016nwo} in higher dimensions using results from \cite{Shaghoulian:2015kta}. This can be utilized to find whether there is a similar suppression in contributions from light primaries for higher dimensional CFTs. We have also relied on the mean value of the structure constant to estimate the light primary contributions to TSD. Hence information about other statistical properties of $c_{\psi \phi \psi}$ will improve our estimations. Finally, our results are only valid for the small subsystem ratio and ideally one would like to find the trace two-norm for all subsystem fraction.

\section*{Acknowledgments}

We would like to thank  Tarun Grover, Tom Hartman, Kenneth Intrilligator, Yunfeng Jiang, Christoph Keller, S.~Prem Kumar, John McGreevy, Tomas Prochazka, Mukund Rangamani, Julian Sonner, Ida Zadeh and Leo Pando Zayas for discussions and comments. We thank Brato Chakrabarti for helping us with the figures.  DD and SP acknowledge the support provided by the U.S. Department of Energy (D.O.E.) under cooperative research agreement DE-SC0009919. The research of SD is supported by the NCCR SwissMAP, funded by the Swiss National Science Foundation.
SD thanks ICTP, Trieste and the Galileo Galilei Institute for Theoretical Physics  (within the program ``New Developments in AdS$_3$/CFT$_2$ Holography") for the hospitality during the completion of this work. In addition, 
SD thanks the INFN and ACRI for partial support through a YITP fellowship.

\appendix

\section{Calculations of $A_k$}\label{app-b}
In this appendix we shall provide details for the calculation of the deviation \eqref{deviation-def}.

\subsection*{$A_T$}
Under the map from the cylinder($w$) to the plane($z$), i.e, $z = e^{-2\pi i w/L}$ the holomorphic part of the stress tensor on the cylinder is given by, 
\beq
T(w=\ell/2 ) = - \frac{ 4 \pi^2}{L^2} e^{-2\pi i \ell/L} T(z=e^{-\pi i \ell/L}) + \frac{ c\pi^2}{6 L^2}. \label{trule}
\eeq
On the plane, the following 3-point function is 
\beq
\bra{0} O_{h,\bar h}(z_1) T(z) O_{h,\bar h}(z_2) \ket{0} = \frac{h ( z_1 -z_2 )^{2-2h}}{(z-z_1)^2 (z-z_2)^2} \frac{1}{(\bar{z}_1-\bar{z}_2)^{2\bar{h}}}.\label{sts-plane}
\eeq
And under the transformation taking the thermal cylinder ($\omega$) to the plane, $z = e^{2\pi \omega/\beta}$, ignoring $e^{-L/\beta}$ corrections, 
\bea
\vev{T(\omega=\ell/2)}_{i\beta/L} &=& - \frac{c\pi^2}{6 \beta^2}.\nnn
\eea
Using (\ref{trule},\ref{sts-plane}) in the cylinder the correlator for the normalized state is,
\bea
&\null& \bra{\psi} T(\ell/2)    \ket{\psi }_{\cyl} \\ &=&  \lim_{z_1,\bar{z}_1 \rightarrow \infty} z_1^{2h} \bar{z}_1^{2\bar{h}} \bra{0} \psi^\dagger(z_1,\bar{z}_1) \bigg(- \frac{ 4 \pi^2}{L^2} e^{-2\pi i \ell/L} T(z=e^{-\pi i \ell/L}) + \frac{ c\pi^2}{6 L^2}\bigg) \psi(0)\ket{0}, \nn
&=& \ \frac{\pi^2 c}{6 L^2} - \frac{4 \pi^2 h}{L^2} . \nnn
\eea
If we include the $e^{-L/\beta}$ corrections, \cite{Chen:2016lbu} in the high temperature regime
\beq
\frac{\vev{T}_{\torus}}{\Tr (e^{-\beta H}) }  = - \frac{ \pi^2 c}{6 \beta^2} + \frac{ 8 \pi^2}{\beta^2} e^{-4\pi L/\beta} + \frac{12\pi^2}{\beta^2} e^{-6\pi L/\beta} + \dots 
\label{Ttorus}
\eeq
Putting together everything the equation for $A_T$ in equation (\ref{Ak}) follows.

\subsection*{$A_\Lambda$} 

Under the map $z \rightarrow z(w ) = e^{-2\pi i w/L }$ from cylinder(w) to plane(z) the level 4 quasi-primary transforms as, 
\begin{align}
&\Lambda(w=\ell/2)  \nn& = \frac{16 \pi^4 e^{-4 \pi i \ell/L } }{L^4} \Lambda(z=e^{-\pi i \ell/L}) - \frac{ 4( 5c + 22 )}{15 L^4} e^{-2\pi i \ell/L} \pi^4 T(z=e^{-\pi i \ell/L}) - \frac{c(5c + 22) \pi^4}{180 L^4}.
\end{align}
After taking our correlator to the plane, in addition to \eqref{sts-plane} if we use,
\beq
\langle\psi(z_1) \Lambda(z) \psi(z_2)\rangle_{\mathbb{C}^2}  = - \frac{5h(z_1 -z_2)^2}{2(z-z_1)^2 (z-z_2)^4} -\frac{5h(z_1 -z_2)^2}{2(z-z_2)^2 (z-z_1)^4} + \frac{ h(6 + 5h) (z_1 -z_2)^4}{5(z-z_1)^4 (z-z_2)^4}.
\eeq
then the equation for $\bra{\psi} \Lambda(w=l/2) \ket{\psi}_{cyl}$ in $A_\Lambda$ follows. Finally, ignoring $e^{-L/\beta}$ corrections, in the high temperature limit we note that in the one point function of $\Lambda$ on the torus only the anomalous piece survives, i.e,
\beq
\vev{\Lambda}_{\torus} =   \frac{c(5c + 22) \pi^4}{180 \beta^4}.
\eeq
Putting things together, $A_\Lambda$ in \eqref{Ak} follows. 

\section{Calculations of $C_{k_1,k_2}$}\label{app-c}
In this appendix, we sketch out the derivation of the coefficients, $C_{k_1,k_2}$, appearing in the short interval expansion. Subsequently, we list out all the relevant coefficients, what we need for our calculation. Following~\cite{Calabrese:2010he}, we use following expression to evaluate $C_{k_1,k_2}$:
\begin{align}\label{cardy}
C_{k_1,k_2}= \frac{Z_2}{Z_{1}^{2}}\ \frac{1}{n_{k_1}n_{k_2}}\ \underset{z_j \rightarrow \infty}{\text{lim}}\ \left(\prod_{j} \left(z_j-\tfrac{\ell}{2}\right)^{2(h_{k_1}+h_{k_2})}\left(\bar{z_j}-\tfrac{\ell}{2}\right)^{2(\bar{h}_{k_1}+\bar{h}_{k_2})}\right)\langle \phi_{k_1}(z_1)\phi_{k_2}(z_2)\rangle_{\text{two sheets}}.
\end{align}
where $n_{k_1},n_{k_2}$ comes due to normalisation of $\phi_{k}$'s,  to be precise, the following holds:
\begin{align}
\langle \phi_{k_1}(w_1)\phi_{k}(w_2)\rangle_{\text{one sheet}} = n_{k_1}\delta_{kk_1} (w_1-w_2)^{-2h_{k_1}} (\bar{w}_1-\bar{w}_2)^{-2\bar{h}_{k_1}}.
\end{align}
Here, $z_{j}$ is the coordinate on the $j$th sheet of a $2$-sheeted manifold, where the operator $\phi_{k_{j}}$ gets inserted, $Z_{2}$ is partition function on a $2$ sheeted plane, while $Z_{1}$ is the partition function on $1$-sheeted plane. We also mention that the appearance of $\frac{\ell}{2}$ is due to the fact that the entanglement cut is on $\left(0,l\right)$ on the 2-sheeted Riemann manifold. The prefactor involving partition function is given by :
\begin{align}
\frac{Z_{2}}{Z_{1}^{2}} \propto \left(\frac{\ell}{\epsilon}\right)^{-\frac{c}{4}}.
\end{align}
 where $\epsilon$ is the UV-regulator

Next we derive the coefficients $C_{T,0},C_{T,T},C_{T,\bar{T}}$ explicitly. For thsi purpose, we adopt the following convention: $z$ is a coordinate on a $2$-sheeted plane while $w$ is a coordinate on a $1$-sheeted plane and the entanglement cut extends along $(0,l)$. The uniformisation map, relating the two co-ordinates is given by
\begin{align}
w(z)=\sqrt{\frac{z}{z-l}},\qquad z(w)= \frac{w^2 \ell}{w^2-1}.
\end{align}
Using the transformation property for $T$ under a mapping $z=f(w)$:
\begin{align}
T(z)= \left(\frac{dz}{dw}\right)^{-2}T(w)+ \frac{c}{12}\{w;z\},
\end{align}
we arrive at
\begin{align}\label{eq:Ttwo}
T(z)=\frac{\left(w^2-1\right)^4}{4 \ell^2 w^2} \left(T(w)+ \frac{c}{8w^2}\right),\qquad \langle T(z)\rangle_{\text{two sheet}} =\frac{c\left(w^2-1\right)^4}{32 \ell^2 w^4}.
\end{align}

To evaluate $C_{T,0}$ the operator $T$ is inserted at $z_1$ (this is the $\phi_{k_1}$ as in \eqref{cardy}) while nothing is inserted at $z_2$ (this amounts inserting identity, which is the $\phi_{k_2}$ as in \eqref{cardy}). So we have, 
\begin{align}
C_{T,0}= \frac{Z_2}{Z_{1}^{2}}\ \frac{2}{c}\ \underset{z_j \rightarrow \infty}{\text{lim}}\ \left(z_1-\frac{\ell}{2}\right)^{4}\langle T(z_1)\rangle_{\text{two sheet}},
\end{align}
which can be recast in terms of $w$ coordinates as
\begin{align}
C_{T,0}= \frac{Z_2}{Z_{1}^{2}}\ \frac{2}{c}\ \underset{w_1 \rightarrow 1}{\text{lim}}\ \frac{\ell^4}{16}\left(\frac{w_1^2+1}{w_1^2-1}\right)^{4}\langle T(z_1)\rangle_{\text{two sheet}}.
\end{align}

Using Eq.~\eqref{eq:Ttwo}, we have
\begin{align}
C_{T,0}=\frac{Z_2}{Z_{1}^{2}} \frac{\ell^{2}}{16}.
\end{align}

Similarly, to evaluate $C_{T,T}$ one $T$ is inserted at $z_1=z(w_1)$ while the other one is inserted at $z_2=z(w_2)$ and 
\begin{align}
C_{T,T}= \frac{Z_2}{Z_{1}^{2}}\ \frac{4}{c^2}\ \underset{w_i \rightarrow \pm1}{\text{lim}}\ \frac{\ell^8}{256}\left(\frac{w_1^2+1}{w_1^2-1}\right)^{4}\left(\frac{w_2^2+1}{w_2^2-1}\right)^{4}\langle T(z(w_1))T(z(w_2))\rangle_{\text{two sheet}},
\end{align}
We use the same uniformisation map and $TT$ OPE on plane to arrive at
\begin{align}
C_{T,T}=\frac{Z_2}{Z_{1}^{2}}\ \frac{4}{c^2}\ \frac{\ell^4}{16}\left(\frac{c}{2^{5}}+\frac{c^{2}}{64}\right)=\frac{Z_2}{Z_{1}^{2}}\ \frac{\ell^{4}}{256}\left(1+\frac{2}{c}\right).
\end{align}

To evaluate $C_{T,\bar{T}}$, we insert $T$ at $z_1$ and $\bar{T}$ at $z_2$, again use the same uniformization map. Here we note that $\langle T\bar{T}\rangle=\langle T\rangle \langle \bar{T}\rangle$, so we have 
\begin{align}
C_{T,\bar{T}}= \frac{Z_2}{Z_{1}^{2}} \frac{\ell^{4}}{256} .
\end{align}

Using similar arguments, one can figure out $C_{\Lambda,\Lambda}, C_{\bar{\Lambda},\bar{\Lambda}},  C_{\Lambda,\bar{\Lambda}}$. All we need to figure out are $\langle\Lambda\Lambda\rangle$, $n_{\Lambda}$  and how $\Lambda$ transforms under a conformal transformations, to be specific, the uniformization map. These are provided below 
\begin{align}
\langle\Lambda(w_1)\Lambda(w_2)\rangle_{\text{one sheet}} = \frac{c (5 c+22)}{10 (w_1-w_2)^8}\ , \qquad n_{\Lambda}=\frac{c (5 c+22)}{10}\ ,
\end{align}
and under uniformization map, it transforms as follows:
\begin{align}
\Lambda(z) = \Lambda(w) \frac{\left(w^2-1\right)^8}{16 \ell^4 w^4}  + \frac{(22+5c)}{320}\frac{\left(w^2-1\right)^8}{ \ell^4 w^6}T(w)+ \frac{c (5 c+22) \left(w^2-1\right)^8}{5120 \ell^4 w^8}.
\end{align}
We list some of the other relevant coefficients calculated in the above way.
\begin{align} 
&C^{\text{primary}}_{k_1,k_2 } = \frac{Z_2}{Z_{1}^{2}}  \bigg( \frac{i\ell}{4} \bigg)^{ 2(h_{k_1 } + \bar{h}_{k_1} )}\delta_{k_1,k_2}\quad \ \ \, \,
C_{T,\bar{T}} = \frac{Z_2}{Z_{1}^{2}} \frac{\ell^{4}}{256} \nn
&C_{T,T}=C_{\bar{T},\bar{T}}=\frac{Z_2}{Z_{1}^{2}}\ \frac{\ell^{4}}{256}\left(1+\frac{2}{c}\right)
\quad\quad\ \  C_{T,\Omega}= C_{\bar{T},\Omega}=  \frac{Z_2}{Z_{1}^{2}}\frac{\ell^6}{4096}\bigg( 1 + \frac{2}{c} \bigg) \label{Ckj} \\
&C_{T,\Lambda}= C_{\bar{T},\bar{\Lambda}} = \frac{Z_2}{Z_1^2} \frac{\ell^6}{8192}\bigg(1 + \frac{4}{c}\bigg)\nonumber \qquad \ \  \  C_{T,\bar{\Lambda}} = C_{\bar{T},\Lambda} = \frac{Z_2}{Z_1^2} \frac{\ell^6}{2^{13}} \nn
&C_{T,\bar{\Lambda}} = C_{\bar{T},\Lambda} = \frac{Z_2}{Z_1^2} \frac{\ell^6}{2^{13}}  \qquad\qquad\qquad\quad \ \ C_{\Omega,\Omega}=\frac{Z_2}{Z_{1}^{2}}\ \frac{\ell^{8}}{256^2}\left(1+\frac{2}{c}\right)^{2} \nn
&C_{\Omega,\Omega}=\frac{Z_2}{Z_{1}^{2}}\ \frac{\ell^{8}}{256^2}\left(1+\frac{2}{c}\right)^{2} \qquad \qquad \quad \, \, C_{\Lambda,\Lambda}=C_{\bar{\Lambda},\bar{\Lambda}}=   \frac{Z_2}{Z_{1}^{2}} \frac{l{^8}(5 c^2+62 c+216)}{262144 c (5 c+22)} \nonumber \\
&C_{\Lambda,\bar{\Lambda}}= \frac{Z_2}{Z_{1}^{2}} \frac{\ell^{8}}{512^{2}} \qquad \qquad\qquad\qquad\qquad   \ \, C_{\Lambda, \Omega} = C_{\bar{ \Lambda},\Omega} =  \frac{Z_2}{Z_{1}^{2}} \frac{\ell^{8}}{131072}\bigg(1 + \frac{4}{c} \bigg)\nn
&C_{{\mathcal{B}},0} = -\frac{Z_2}{Z_1^2} \frac{ 8(61+35c) - 93}{73728(29 + 70c)} \ell^6 \qquad \quad \ \ \,
C_{{\mathcal{D}},0} = \frac{Z_2}{Z_1^2} \frac{ \ell^6}{24576}. \nnn
\end{align}
where $\Lambda=\,  \normord{TT}-\frac{3}{10}\partial^{2}T$, $\mathcal{B} = \normord{\partial T \partial T} - \frac{4}{5} \normord{ \partial^2 T T } - \frac{1}{42} \partial^4 T$ and $\mathcal{D} = \normord{ T(\normord{TT})} - \frac{9}{10} \normord{\partial^2 T T} -\frac{1}{28} \partial^4 T + \frac{93}{70c + 29} \mathcal{B}$. We have checked that our answers match with \cite{Chen:2013dxa}. 

\section{Second Renyi entropies and additional checks}\label{app-d}
Consider a 1+1d CFT living on a circle of circumference, $L$. The coefficients $C_{k_i,k_j}$ calculated above can be applied to the short interval expansion of the $\tr(\rho_A^2)$, where $\rho_{A}$ is the reduced density corresponding to an entanglement cut extending from $0$ to $\ell$ and time $t=0$. This gives the exponential of the second Renyi entropy. In the case when the CFT is in a thermal state with inverse temperature $\beta$, in the high temperature regime, $\beta/L \rightarrow 0$, the quantity, $\tr(\rho_A^\beta)^2$ is exactly known as the two-point function of $\mathbb{Z}_2$ twist fields. This is given by:
\beq
\Tr(\rho_A^\beta)^2 = \la \sigma_2 (0) \bar \sigma_2 (\ell)  \ra_\torus  = {1 \over \big|\tfrac{\beta}{\pi} \sinh(\tfrac{\pi \ell}{\beta}) \big|^{c/4}}.
\eeq
Consequently, the short interval expansion i.e expansion of the above, in powers of $\frac{\ell}{\beta}$, is given by
\bea
\label{r2-sie}
\Tr(\rho_A^\beta)^2 = \bigg(\frac{\ell}{\epsilon}\bigg)^{-c/4} \bigg[1 - \tfrac{c\pi^2}{24} \left(\tfrac{\ell}{\beta}\right)^2 + \tfrac{\pi ^4 c (5 c+8)}{5760} \left(\tfrac{\ell}{\beta}\right)^4 - \tfrac{\pi ^6 c \left(35 c^2+168 c+256\right)}{2903040} \left(\tfrac{\ell}{\beta}\right)^6 + \cdots \bigg]
\eea
On the other hand, the sewing procedure, as done in the main text to evaluate trace two norm, leads to following short interval expansion:
\bea\label{tracecheck}
\Tr(\rho_A^\beta)^2 &=&\bigg(\frac{\ell}{\epsilon}\bigg)^{-c/4}  \bigg[   1 + \left(2\hat{C}_{T,0}\vev{T}_\torus + 2\hat{C}_{\bar{T},0}\vev{\bar T}_\torus \right) + \bigg(2 \hat{C}_{\Lambda,0}\vev{\Lambda}_\torus + 2\hat{C}_{\bar{\Lambda},0}\vev{\bar \Lambda}_\torus  \nn
&+&2\hat{C}_{\Omega,0}\vev{ \Omega}_\torus+  \hat{C}_{T,T} \vev{T}^2_\torus + \hat{C}_{\bar T,\bar T} \vev{\bar T}^2_\torus + 2\hat{C}_{T,\bar T}\vev{T}_\torus \vev{\bar T}_\torus\bigg) + \bigg( 2\hat{C}_{T, \Lambda}\vev{\Lambda}_\torus \vev{T}_\torus  \nn
&+&2\hat{C}_{\bar{T}, \bar \Lambda}\vev{\bar \Lambda}_\torus \vev{\bar T}_\torus + 2\hat{C}_{\bar{T}, \Lambda}\vev{\Lambda}_\torus \vev{\bar T}_\torus+2\hat{C}_{T, \bar \Lambda}\vev{\bar \Lambda}_\torus \vev{T}_\torus +2\hat{C}_{T, \Omega}\vev{\Omega}_\torus \vev{T}_\torus 
  \nn
&+& 2\hat{C}_{\bar T, \Omega}\vev{\Omega}_\torus \vev{\bar T}_\torus + 2\hat{C}_{ T\bar{\Lambda},0}\vev{T}_\torus \vev{\bar \Lambda}_\torus +
2\hat{C}_{ \bar{T}\Lambda,0}\vev{\bar{T}}_\torus \vev{ \Lambda}_\torus
+  2\hat{C}_{0, \mathcal{B}}\vev{\mathcal{B}}_\torus \nn
&+&2 \hat{C}_{0, \mathcal{D}} \vev{\mathcal{D}}_\torus\bigg)  +\dots     \bigg].
\eea
where  $\left(\frac{\ell}{\epsilon}\right)^{-c/4}\hat{C}_{k_1,k_2}=C_{k_1,k_2}$ and $\hat{C}_{0,0}=1$. 
We have explicitly checked that on substituting the values of $C_{k_i,k_j}$ and the one-point functions of the quasi-primaries above, the short interval expansion \eqref{r2-sie} gets reproduced. (No primaries $\phi_{k}$ contribute to \eqref{tracecheck} as one point function of primaries on a cylinder, $\vev{\phi_k}$ is identically $0$.)

Similarly for the finite energy density eigenstate, $\ket{\psi}$, we can use the short interval expansion to calculate the universal contribution to the trace of the reduced density matrix squared, 
\bea
\label{renyi2psi}
\Tr( \rho_A^\psi)^2 &=& \bigg(\frac{\ell}{\epsilon}\bigg)^{-c/4}  \bigg[   1 + \left(2\hat{C}_{T,0}\vev{T}_\psi + 2\hat{C}_{\bar{T},0}\vev{\bar T}_\psi \right)  + \bigg(2 \hat{C}_{\Lambda,0} \vev{\Lambda}_\psi +2\hat{C}_{\bar{\Lambda},0}\vev{\bar \Lambda}_\psi \nn
&+&  2\hat{C}_{\Omega,0}\vev{ \Omega}_\psi
    +  \hat{C}_{T,T} \vev{T}^2_\psi + \hat{C}_{\bar T,\bar T} \vev{\bar T}^2_\psi 
+ 2\hat{C}_{T,\bar T}\vev{T}_\psi \vev{\bar T}_\psi\bigg) + \dots  \nn
   &=&  \bigg(\frac{\ell}{\epsilon}\bigg)^{-c/4}  \bigg[   1 - \tfrac{ \pi^2(24 h - c)}{24} \left(\tfrac{\ell}{L} \right)^2 + \tfrac{ \pi^4 (2 + 3c)(24h-c)^2}{4608 c} \left( \tfrac{\ell}{L}\right)^4 + \dots \bigg]. 
\eea
where, $\vev{O}_\psi = \vev{\psi| O |\psi}_\cyl$ and we have kept terms till quartic order in subsystem fraction. Thus till this order, the second Renyi entropy is given by, $S_2 = - \log \Tr(\rho_A^\psi)^2$, 
\beq
S_2 = \frac{c}{4} \log \left( \tfrac{\ell}{\epsilon} \right) + \tfrac{ \pi^2(24 h - c)}{24} \left(\tfrac{\ell}{L} \right)^2 + \tfrac{ \pi^4 (c-2 )(24h-c)^2}{4608 c} \left( \tfrac{\ell}{L}\right)^4 + \mathrm{O}\left( \left(\tfrac{\ell}{L}\right)^6 \right).
\eeq

\bibliographystyle{bibstyle}
\bibliography{collection}

\providecommand{\href}[2]{#2}\begingroup\begin{thebibliography}{10}

\bibitem{Deutsch:1991}
J.~M. Deutsch, {\it Quantum statistical mechanics in a closed system},
  \href{http://dx.doi.org/10.1103/PhysRevA.43.2046}{{\em Phys. Rev. A}
  {\bfseries 43} (Feb, 1991) 2046--2049}.
  \url{https://link.aps.org/doi/10.1103/PhysRevA.43.2046}.

\bibitem{Srednicki:1994}
M.~{Srednicki}, {\it {Chaos and quantum thermalization}},
  \href{http://dx.doi.org/10.1103/PhysRevE.50.888}{{\em Phys. Rev. E}
  {\bfseries 50} (Aug., 1994) 888--901},
  \href{http://arxiv.org/abs/cond-mat/9403051}{{\ttfamily cond-mat/9403051}}.

\bibitem{rigol2008thermalization}
M.~Rigol, V.~Dunjko, and M.~Olshanii, {\it Thermalization and its mechanism for
  generic isolated quantum systems},  {\em Nature} {\bfseries 452} no.~7189,
  (2008) 854--858.

\bibitem{popescu2006entanglement}
S.~Popescu, A.~J. Short, and A.~Winter, {\it Entanglement and the foundations
  of statistical mechanics},  {\em Nature Physics} {\bfseries 2} no.~11, (2006)
  754--758.

\bibitem{nandkishore2015many}
R.~Nandkishore and D.~A. Huse, {\it Many-body localization and thermalization
  in quantum statistical mechanics},  {\em Annu. Rev. Condens. Matter Phys.}
  {\bfseries 6} no.~1, (2015) 15--38.

\bibitem{huse2014}
H.~Kim, T.~N. Ikeda, and D.~A. Huse, {\it Testing whether all eigenstates obey
  the eigenstate thermalization hypothesis},  {\em Physical Review E}
  {\bfseries 90} no.~5, (2014) 052105.

\bibitem{Garrison:2015lva}
J.~R. Garrison and T.~Grover, {\it {Does a single eigenstate encode the full
  Hamiltonian?}},
\href{http://arxiv.org/abs/1503.00729}{{\ttfamily arXiv:1503.00729
  [cond-mat.str-el]}}.

\bibitem{alba2015eigenstate}
V.~Alba, {\it Eigenstate thermalization hypothesis and integrability in quantum
  spin chains},  {\em Physical Review B} {\bfseries 91} no.~15, (2015) 155123.

\bibitem{Nandy:2016fwv}
S.~Nandy, A.~Sen, A.~Das, and A.~Dhar, {\it {Eigenstate Gibbs Ensemble in
  Integrable Quantum Systems}},
  \href{http://dx.doi.org/10.1103/PhysRevB.94.245131}{{\em Phys. Rev.}
  {\bfseries B94} no.~24, (2016) 245131},
\href{http://arxiv.org/abs/1605.09225}{{\ttfamily arXiv:1605.09225
  [cond-mat.stat-mech]}}.

\bibitem{berry1977regular}
M.~V. Berry, {\it Regular and irregular semiclassical wavefunctions},  {\em
  Journal of Physics A: Mathematical and General} {\bfseries 10} no.~12, (1977)
  2083.

\bibitem{Bernard:2016nci}
D.~Bernard and B.~Doyon, {\it {Conformal field theory out of equilibrium: a
  review}},  \href{http://dx.doi.org/10.1088/1742-5468/2016/06/064005}{{\em J.
  Stat. Mech.} {\bfseries 1606} no.~6, (2016) 064005},
\href{http://arxiv.org/abs/1603.07765}{{\ttfamily arXiv:1603.07765
  [cond-mat.stat-mech]}}.

\bibitem{Lashkari:2016vgj}
N.~Lashkari, A.~Dymarsky, and H.~Liu, {\it {Eigenstate Thermalization
  Hypothesis in Conformal Field Theory}},
\href{http://arxiv.org/abs/1610.00302}{{\ttfamily arXiv:1610.00302 [hep-th]}}.

\bibitem{Cardy:2014rqa}
J.~Cardy, {\it {Thermalization and Revivals after a Quantum Quench in Conformal
  Field Theory}},  \href{http://dx.doi.org/10.1103/PhysRevLett.112.220401}{{\em
  Phys. Rev. Lett.} {\bfseries 112} (2014) 220401},
\href{http://arxiv.org/abs/1403.3040}{{\ttfamily arXiv:1403.3040
  [cond-mat.stat-mech]}}.

\bibitem{Mandal:2015jla}
G.~Mandal, R.~Sinha, and N.~Sorokhaibam, {\it {Thermalization with chemical
  potentials, and higher spin black holes}},
  \href{http://dx.doi.org/10.1007/JHEP08(2015)013}{{\em JHEP} {\bfseries 08}
  (2015) 013},
\href{http://arxiv.org/abs/1501.04580}{{\ttfamily arXiv:1501.04580 [hep-th]}}.

\bibitem{Caputa:2014eta}
P.~Caputa, J.~Simón, A.~Stikonas, and T.~Takayanagi, {\it {Quantum Entanglement
  of Localized Excited States at Finite Temperature}},
  \href{http://dx.doi.org/10.1007/JHEP01(2015)102}{{\em JHEP} {\bfseries 01}
  (2015) 102},
\href{http://arxiv.org/abs/1410.2287}{{\ttfamily arXiv:1410.2287 [hep-th]}}.

\bibitem{Asplund:2014coa}
C.~T. Asplund, A.~Bernamonti, F.~Galli, and T.~Hartman, {\it {Holographic
  Entanglement Entropy from 2d CFT: Heavy States and Local Quenches}},
  \href{http://dx.doi.org/10.1007/JHEP02(2015)171}{{\em JHEP} {\bfseries 02}
  (2015) 171},
\href{http://arxiv.org/abs/1410.1392}{{\ttfamily arXiv:1410.1392 [hep-th]}}.

\bibitem{Fitzpatrick:2015foa}
A.~L. Fitzpatrick, J.~Kaplan, M.~T. Walters, and J.~Wang, {\it {Hawking from
  Catalan}},  \href{http://dx.doi.org/10.1007/JHEP05(2016)069}{{\em JHEP}
  {\bfseries 05} (2016) 069},
\href{http://arxiv.org/abs/1510.00014}{{\ttfamily arXiv:1510.00014 [hep-th]}}.

\bibitem{Fitzpatrick:2015dlt}
A.~L. Fitzpatrick and J.~Kaplan, {\it {Conformal Blocks Beyond the
  Semi-Classical Limit}},
  \href{http://dx.doi.org/10.1007/JHEP05(2016)075}{{\em JHEP} {\bfseries 05}
  (2016) 075},
\href{http://arxiv.org/abs/1512.03052}{{\ttfamily arXiv:1512.03052 [hep-th]}}.

\bibitem{Fitzpatrick:2015zha}
A.~L. Fitzpatrick, J.~Kaplan, and M.~T. Walters, {\it {Virasoro Conformal
  Blocks and Thermality from Classical Background Fields}},
  \href{http://dx.doi.org/10.1007/JHEP11(2015)200}{{\em JHEP} {\bfseries 11}
  (2015) 200},
\href{http://arxiv.org/abs/1501.05315}{{\ttfamily arXiv:1501.05315 [hep-th]}}.

\bibitem{Dymarsky:2016aqv}
A.~Dymarsky, N.~Lashkari, and H.~Liu, {\it {Subsystem ETH}},
\href{http://arxiv.org/abs/1611.08764}{{\ttfamily arXiv:1611.08764
  [cond-mat.stat-mech]}}.

\bibitem{Anous:2016kss}
T.~Anous, T.~Hartman, A.~Rovai, and J.~Sonner, {\it {Black Hole Collapse in the
  1/c Expansion}},  \href{http://dx.doi.org/10.1007/JHEP07(2016)123}{{\em JHEP}
  {\bfseries 07} (2016) 123},
\href{http://arxiv.org/abs/1603.04856}{{\ttfamily arXiv:1603.04856 [hep-th]}}.

\bibitem{He:2017vyf}
S.~He, F.-L. Lin, and J.-j. Zhang, {\it {Subsystem eigenstate thermalization
  hypothesis for entanglement entropy in CFT}},
\href{http://arxiv.org/abs/1703.08724}{{\ttfamily arXiv:1703.08724 [hep-th]}}.

\bibitem{Sarosi:2016oks}
G.~Sarosi and T.~Ugajin, {\it {Relative entropy of excited states in two
  dimensional conformal field theories}},
  \href{http://dx.doi.org/10.1007/JHEP07(2016)114}{{\em JHEP} {\bfseries 07}
  (2016) 114},
\href{http://arxiv.org/abs/1603.03057}{{\ttfamily arXiv:1603.03057 [hep-th]}}.

\bibitem{CARDY1986186}
J.~L. Cardy, {\it Operator content of two-dimensional conformally invariant
  theories},
  \href{http://dx.doi.org/http://dx.doi.org/10.1016/0550-3213(86)90552-3}{{\em
  Nuclear Physics B} {\bfseries 270} (1986) 186 -- 204}.
  \url{http://www.sciencedirect.com/science/article/pii/0550321386905523}.

\bibitem{Cardy:2007mb}
J.~L. Cardy, O.~A. Castro-Alvaredo, and B.~Doyon, {\it {Form factors of
  branch-point twist fields in quantum integrable models and entanglement
  entropy}},  \href{http://dx.doi.org/10.1007/s10955-007-9422-x}{{\em J.
  Statist. Phys.} {\bfseries 130} (2008) 129--168},
\href{http://arxiv.org/abs/0706.3384}{{\ttfamily arXiv:0706.3384 [hep-th]}}.

\bibitem{Calabrese:2009qy}
P.~Calabrese and J.~Cardy, {\it {Entanglement entropy and conformal field
  theory}},  \href{http://dx.doi.org/10.1088/1751-8113/42/50/504005}{{\em
  J.Phys.} {\bfseries A42} (2009) 504005},
\href{http://arxiv.org/abs/0905.4013}{{\ttfamily arXiv:0905.4013
  [cond-mat.stat-mech]}}.

\bibitem{Headrick:2010zt}
M.~Headrick, {\it {Entanglement Renyi entropies in holographic theories}},
  \href{http://dx.doi.org/10.1103/PhysRevD.82.126010}{{\em Phys. Rev.}
  {\bfseries D82} (2010) 126010},
\href{http://arxiv.org/abs/1006.0047}{{\ttfamily arXiv:1006.0047 [hep-th]}}.

\bibitem{Sen:1990bt}
A.~Sen, {\it {Some aspects of conformal field theories on the plane and higher
  genus Riemann surfaces}},
\href{http://dx.doi.org/10.1007/BF02846591}{{\em Pramana} {\bfseries 35} (1990)
  205--286}.

\bibitem{Gaberdiel:2010jf}
M.~R. Gaberdiel, C.~A. Keller, and R.~Volpato, {\it {Genus Two Partition
  Functions of Chiral Conformal Field Theories}},
  \href{http://dx.doi.org/10.4310/CNTP.2010.v4.n2.a2}{{\em Commun. Num. Theor.
  Phys.} {\bfseries 4} (2010) 295--364},
\href{http://arxiv.org/abs/1002.3371}{{\ttfamily arXiv:1002.3371 [hep-th]}}.

\bibitem{Headrick:2015gba}
M.~Headrick, A.~Maloney, E.~Perlmutter, and I.~G. Zadeh, {\it {Rényi entropies,
  the analytic bootstrap, and 3D quantum gravity at higher genus}},
  \href{http://dx.doi.org/10.1007/JHEP07(2015)059}{{\em JHEP} {\bfseries 07}
  (2015) 059},
\href{http://arxiv.org/abs/1503.07111}{{\ttfamily arXiv:1503.07111 [hep-th]}}.

\bibitem{Kraus:2016nwo}
P.~Kraus and A.~Maloney, {\it {A Cardy Formula for Three-Point Coefficients:
  How the Black Hole Got its Spots}},
\href{http://arxiv.org/abs/1608.03284}{{\ttfamily arXiv:1608.03284 [hep-th]}}.

\bibitem{Bazhanov:1994ft}
V.~V. Bazhanov, S.~L. Lukyanov, and A.~B. Zamolodchikov, {\it {Integrable
  structure of conformal field theory, quantum KdV theory and thermodynamic
  Bethe ansatz}},  \href{http://dx.doi.org/10.1007/BF02101898}{{\em Commun.
  Math. Phys.} {\bfseries 177} (1996) 381--398},
\href{http://arxiv.org/abs/hep-th/9412229}{{\ttfamily arXiv:hep-th/9412229
  [hep-th]}}.

\bibitem{Sasaki:1987mm}
R.~Sasaki and I.~Yamanaka, {\it {Virasoro Algebra, Vertex Operators, Quantum
  {Sine-Gordon} and Solvable Quantum Field Theories}},
{\em Adv. Stud. Pure Math.} {\bfseries 16} (1988) 271--296.

\bibitem{deBoer:2016bov}
J.~de~Boer and D.~Engelhardt, {\it {Remarks on thermalization in 2D CFT}},
  \href{http://dx.doi.org/10.1103/PhysRevD.94.126019}{{\em Phys. Rev.}
  {\bfseries D94} no.~12, (2016) 126019},
\href{http://arxiv.org/abs/1604.05327}{{\ttfamily arXiv:1604.05327 [hep-th]}}.

\bibitem{Cardy:2015xaa}
J.~Cardy, {\it {Quantum Quenches to a Critical Point in One Dimension: some
  further results}},
  \href{http://dx.doi.org/10.1088/1742-5468/2016/02/023103}{{\em J. Stat.
  Mech.} {\bfseries 1602} no.~2, (2016) 023103},
\href{http://arxiv.org/abs/1507.07266}{{\ttfamily arXiv:1507.07266
  [cond-mat.stat-mech]}}.

\bibitem{Sotiriadis:2015kfa}
S.~Sotiriadis and G.~Martelloni, {\it {Equilibration and GGE in
  interacting-to-free quantum quenches in dimensions $d > 1$}},
  \href{http://dx.doi.org/10.1088/1751-8113/49/9/095002}{{\em J. Phys.}
  {\bfseries A49} no.~9, (2016) 095002},
\href{http://arxiv.org/abs/1505.08150}{{\ttfamily arXiv:1505.08150 [hep-th]}}.

\bibitem{Faulkner:2013yia}
T.~Faulkner, {\it {The Entanglement Renyi Entropies of Disjoint Intervals in
  AdS/CFT}},
\href{http://arxiv.org/abs/1303.7221}{{\ttfamily arXiv:1303.7221 [hep-th]}}.

\bibitem{Barrella:2013wja}
T.~Barrella, X.~Dong, S.~A. Hartnoll, and V.~L. Martin, {\it {Holographic
  entanglement beyond classical gravity}},
  \href{http://dx.doi.org/10.1007/JHEP09(2013)109}{{\em JHEP} {\bfseries 1309}
  (2013) 109},
\href{http://arxiv.org/abs/1306.4682}{{\ttfamily arXiv:1306.4682 [hep-th]}}.

\bibitem{Datta:2013hba}
S.~Datta and J.~R. David, {\it {Renyi entropies of free bosons on the torus and
  holography}},  \href{http://dx.doi.org/10.1007/JHEP04(2014)081}{{\em JHEP}
  {\bfseries 1404} (2014) 081},
\href{http://arxiv.org/abs/1311.1218}{{\ttfamily arXiv:1311.1218 [hep-th]}}.

\bibitem{Krasnov:2000zq}
K.~Krasnov, {\it {Holography and Riemann surfaces}},  {\em Adv. Theor. Math.
  Phys.} {\bfseries 4} (2000) 929--979,
\href{http://arxiv.org/abs/hep-th/0005106}{{\ttfamily arXiv:hep-th/0005106
  [hep-th]}}.

\bibitem{Gubser:1998bc}
S.~S. Gubser, I.~R. Klebanov, and A.~M. Polyakov, {\it {Gauge theory
  correlators from noncritical string theory}},
  \href{http://dx.doi.org/10.1016/S0370-2693(98)00377-3}{{\em Phys. Lett.}
  {\bfseries B428} (1998) 105--114},
\href{http://arxiv.org/abs/hep-th/9802109}{{\ttfamily arXiv:hep-th/9802109
  [hep-th]}}.

\bibitem{Witten:1998qj}
E.~Witten, {\it {Anti-de Sitter space and holography}},  {\em Adv. Theor. Math.
  Phys.} {\bfseries 2} (1998) 253--291,
\href{http://arxiv.org/abs/hep-th/9802150}{{\ttfamily arXiv:hep-th/9802150
  [hep-th]}}.

\bibitem{Bagchi:2015wna}
A.~Bagchi, D.~Grumiller, and W.~Merbis, {\it {Stress tensor correlators in
  three-dimensional gravity}},
  \href{http://dx.doi.org/10.1103/PhysRevD.93.061502}{{\em Phys. Rev.}
  {\bfseries D93} no.~6, (2016) 061502},
\href{http://arxiv.org/abs/1507.05620}{{\ttfamily arXiv:1507.05620 [hep-th]}}.

\bibitem{Shaghoulian:2015kta}
E.~Shaghoulian, {\it {Modular forms and a generalized Cardy formula in higher
  dimensions}},  \href{http://dx.doi.org/10.1103/PhysRevD.93.126005}{{\em Phys.
  Rev.} {\bfseries D93} no.~12, (2016) 126005},
\href{http://arxiv.org/abs/1508.02728}{{\ttfamily arXiv:1508.02728 [hep-th]}}.

\bibitem{Chen:2016lbu}
B.~Chen, J.-B. Wu, and J.-j. Zhang, {\it {Short interval expansion of Rényi
  entropy on torus}},  \href{http://dx.doi.org/10.1007/JHEP08(2016)130}{{\em
  JHEP} {\bfseries 08} (2016) 130},
\href{http://arxiv.org/abs/1606.05444}{{\ttfamily arXiv:1606.05444 [hep-th]}}.

\bibitem{Calabrese:2010he}
P.~Calabrese, J.~Cardy, and E.~Tonni, {\it {Entanglement entropy of two
  disjoint intervals in conformal field theory II}},
  \href{http://dx.doi.org/10.1088/1742-5468/2011/01/P01021}{{\em J. Stat.
  Mech.} {\bfseries 1101} (2011) P01021},
\href{http://arxiv.org/abs/1011.5482}{{\ttfamily arXiv:1011.5482 [hep-th]}}.

\bibitem{Chen:2013dxa}
B.~Chen, J.~Long, and J.-j. Zhang, {\it {Holographic Rényi entropy for CFT with
  W symmetry}},  \href{http://dx.doi.org/10.1007/JHEP04(2014)041}{{\em JHEP}
  {\bfseries 04} (2014) 041},
\href{http://arxiv.org/abs/1312.5510}{{\ttfamily arXiv:1312.5510 [hep-th]}}.

\end{thebibliography}\endgroup

\end{document}